\documentclass[fleqn,usenatbib]{mnras}

\usepackage{newtxtext,newtxmath}

\usepackage[T1]{fontenc}

\DeclareRobustCommand{\VAN}[3]{#2}
\let\VANthebibliography\thebibliography
\def\thebibliography{\DeclareRobustCommand{\VAN}[3]{##3}\VANthebibliography}

\usepackage{graphicx}	
\usepackage{amsmath}	
\usepackage{float}

\title[The ASKAP-EMU ESP LMC Radio Continuum Survey]{The ASKAP-EMU Early Science Project: 888 MHz Radio Continuum Survey of the Large Magellanic Cloud}

\author[Clara M. Pennock et al.]{
Clara M. Pennock,$^{1}$\thanks{E-mail: c.m.pennock@keele.ac.uk}
Jacco Th. van Loon,$^{1}$
Miroslav D. Filipovi\'c,$^{2}$
Heinz Andernach,$^{3}$
\newauthor Frank Haberl,$^{4}$
Roland Kothes,$^{5}$
Emil Lenc,$^{6}$
Lawrence Rudnick,$^{7}$
Sarah V. White,$^{8}$
\newauthor Claudia Agliozzo,$^{9}$
Sonia Ant\'{o}n,$^{10}$
Ivan Boji\v{c}i\'{c},$^{2}$
Dominik J. Bomans,$^{11,12}$
\newauthor Jordan D. Collier,$^{2,6,13}$
Evan J. Crawford,$^{2}$
Andrew M. Hopkins,$^{14}$
\newauthor Kanapathippillai Jeganathan,$^{15}$ 
Patrick J. Kavanagh,$^{16}$ 
B\"arbel S. Koribalski,$^{2,6}$
\newauthor Denis Leahy,$^{17}$
Pierre Maggi,$^{18}$ 
Chandreyee Maitra,$^{4}$
Josh Marvil,$^{19}$
\newauthor Micha\l{} J. Micha\l{}owski,$^{120}$
Ray P. Norris,$^{2,6}$
Joana M. Oliveira,$^{1}$
Jeffrey L. Payne,$^{2}$ 
\newauthor Hidetoshi Sano,$^{21,22,23}$ 
Manami Sasaki,$^{24}$ 
Lister Staveley-Smith,$^{25,26}$ 
Eleni Vardoulaki$^{27}$
\\
$^{1}$Lennard-Jones Laboratories, Keele University, ST5 5BG, UK\\
Affiliations are listed at the end of the paper\\
}

\date{Accepted XXX. Received YYY; in original form ZZZ}

\pubyear{2021}

\begin{document}
\label{firstpage}
\pagerange{\pageref{firstpage}--\pageref{lastpage}}
\maketitle

\begin{abstract}
We present an analysis of a new 120 deg$^{2}$ radio continuum image of the Large Magellanic Cloud (LMC) at 888 MHz with a bandwidth of 288 MHz and beam size of $13\rlap{.}^{\prime\prime}9\times12\rlap{.}^{\prime\prime}1$, from the Australian Square Kilometre Array Pathfinder (ASKAP) processed as part of the Evolutionary Map of the Universe (EMU) survey. The median Root Mean Squared noise is 58 $\mu$Jy beam$^{-1}$. We present a catalogue of 54,612 sources, divided over a G\textsc{old} list (30,866 sources) complete down to 0.5 mJy uniformly across the field, a S\textsc{ilver} list (22,080 sources) reaching down to $<$ 0.2 mJy and a B\textsc{ronze} list (1,666 sources) of visually inspected sources in areas of high noise and/or near bright complex emission. We discuss detections of planetary nebul\ae\ and their radio luminosity function, young stellar objects showing a correlation between radio luminosity and gas temperature, nov\ae\ and X-ray binaries in the LMC, and active stars in the Galactic foreground that may become a significant population below this flux level. We present examples of diffuse emission in the LMC (H\,{\sc ii} regions, supernova remnants, bubbles) and distant galaxies showcasing spectacular interaction between jets and intracluster medium. Among 14,333 infrared counterparts of the predominantly background radio source population we find that star-forming galaxies become more prominent
below 3 mJy compared to active galactic nuclei. We combine the new 888 MHz data with archival Australia Telescope Compact Array data at 1.4 GHz to determine spectral indices; the vast majority display synchrotron emission but flatter spectra occur too. We argue that the most extreme spectral index values are due to variability.
\end{abstract}




\section{Introduction}



The Large Magellanic Cloud (LMC) is a disrupted barred spiral galaxy interacting with the Milky Way and its nearest neighbour, the Small Magellanic Cloud. The Magellanic Stream is excellent evidence for this interaction \citep{Mathewson1974, Murai1980}.  At a distance of $\sim$ 50 kpc \citep{Pietrzynski2013,Pietrzynski2019,Riess2019}, the gas-rich LMC's close proximity and size allows an exquisitely detailed study of  stellar populations and
the interstellar medium (ISM) across an entire galaxy,
providing a comprehensive picture of star formation, stellar feedback and
galaxy structure. It has been a benchmark for astrophysical discoveries, such as the Cepheid period--luminosity relation by Henrietta Leavitt \citep[see][]{Leavitt1908,Leavitt1912} and the best-studied supernova,  SN\,1987A. 

At radio frequencies, we typically observe free--free emission from ionized gas in H\,{\sc ii} regions and planetary nebul\ae\ (PNe) and synchrotron emission from supernova remnants (SNRs), as well as from active galactic nuclei (AGN) and star-forming galaxies in the background. The  LMC is located  away from the Galactic Plane and Galactic Centre, reducing source confusion in the radio band and extinction at UV/optical/near-IR wavelengths.  It has been observed multiple times over the years by telescopes such as MOST \citep[Molongo Observatory Synthesis Telescope; e.g., ][]{Large1981,Mills1985,Mauch2003}, ATCA \citep[Australia Telescope Compact Array; e.g., ][]{Kim1998,Dickel2005,Hughes2007, Murphy2010}, MWA \citep[Murchison Widefield Array; e.g., ][]{For2018} and Parkes \citep{Griffith1993,Filipovic1995,Filipovic1996,Filipovic1998b,Filipovic1998a,Filipovic1998c,Filipovic1998d,Kim2003}. A new generation of radio telescopes, including the Australian Square Kilometre Array Pathfinder \citep[ASKAP;][]{Johnston2008,Hotan2021}  can improve upon the resolution, sensitivity and speed of these observations. ASKAP studies of the LMC and Small Magellanic Cloud \citep[SMC,][]{Joseph2019} can provide greater details on SNRs, PNe, (super)bubbles and their environments, young stellar objects (YSOs), symbiotic (accreting compact object) binaries, AGN and star-forming galaxies.

The LMC has been the subject of extensive multi-wavelength surveys. These include all-sky surveys such as WISE \citep[Wide-field Infrared Survey Explorer;][]{Wright2010} and Gaia \citep{Gaia2018}, as well as Magellanic Cloud specific surveys such as SAGE \citep[Surveying the Agents of a Galaxy's Evolution;][]{Meixner2006}, HERITAGE \citep[HERschel Inventory of the Agents of Galaxy Evolution;][]{Meixner2010}, VMC \citep[VISTA Magellanic Cloud;][]{Cioni2011, Cioni2017}, SMASH \citep[Survey of the Magellanic Stellar History;][]{Nidever2017} and sections of the LMC surveyed by XMM--{\it Newton} \citep{Haberl2014}. 

At radio frequencies AGN and galaxies dominate the source counts. Hence, multi-wavelength surveys from X-ray to far-infrared (far-IR) combined with the new EMU ASKAP data provide a prime opportunity to study a large population of high redshift AGN in the direction of the LMC. These AGN, once found, while interesting in their own right, can be used as a reference frame for proper-motion studies of the LMC \citep[e.g., ][]{Kalli2006,Costa2009,Pedreros2011,vanderMarel2014,Vieira2014,Schmidt2020} and to probe the interstellar medium through the LMC \citep[e.g., ][]{Haberl2001}. The ASKAP's unprecedented sensitivity and resolution allows us to detect and resolve stellar and diffuse sources within the LMC and foreground allowing more systematic radio investigations of populations of such objects ahead of the next generation Square Kilometre Array.   

The LMC was selected as a prime target for the Early Science Project (ESP) as part of the ASKAP \citep{Johnston2008,Hotan2021} commissioning and early science verification. ASKAP is a radio interferometer that allows us to survey the LMC with a sensitivity down to the $\mu$Jy range and a large field of view of 30~deg$^{2}$ \citep{Murphy2013, Leahy2019}. Here we present a new source catalogue from the ASKAP ESP for radio point sources observed towards the LMC, obtained from a mosaic of images taken at 888 MHz ($\lambda = 34$ cm). 

This paper is laid out as follows: Section 2 describes the data used to create the source lists. In section 3 we describe the creation of the source catalogue and compare our work to previous catalogues of point sources towards the LMC and SMC. In section 4 we discuss the Magellanic and Galactic foreground sources. Lastly, in section 5 we discuss the sources seen beyond the LMC and in section 6 we determine and discuss the spectral indices of sources in the new catalogue. The source catalogue and 888 MHz image are made available publicly. 

\section{Data, observing and processing}

\begin{figure}
\centering
    \begin{tabular}{c}
	\includegraphics[width=\columnwidth]{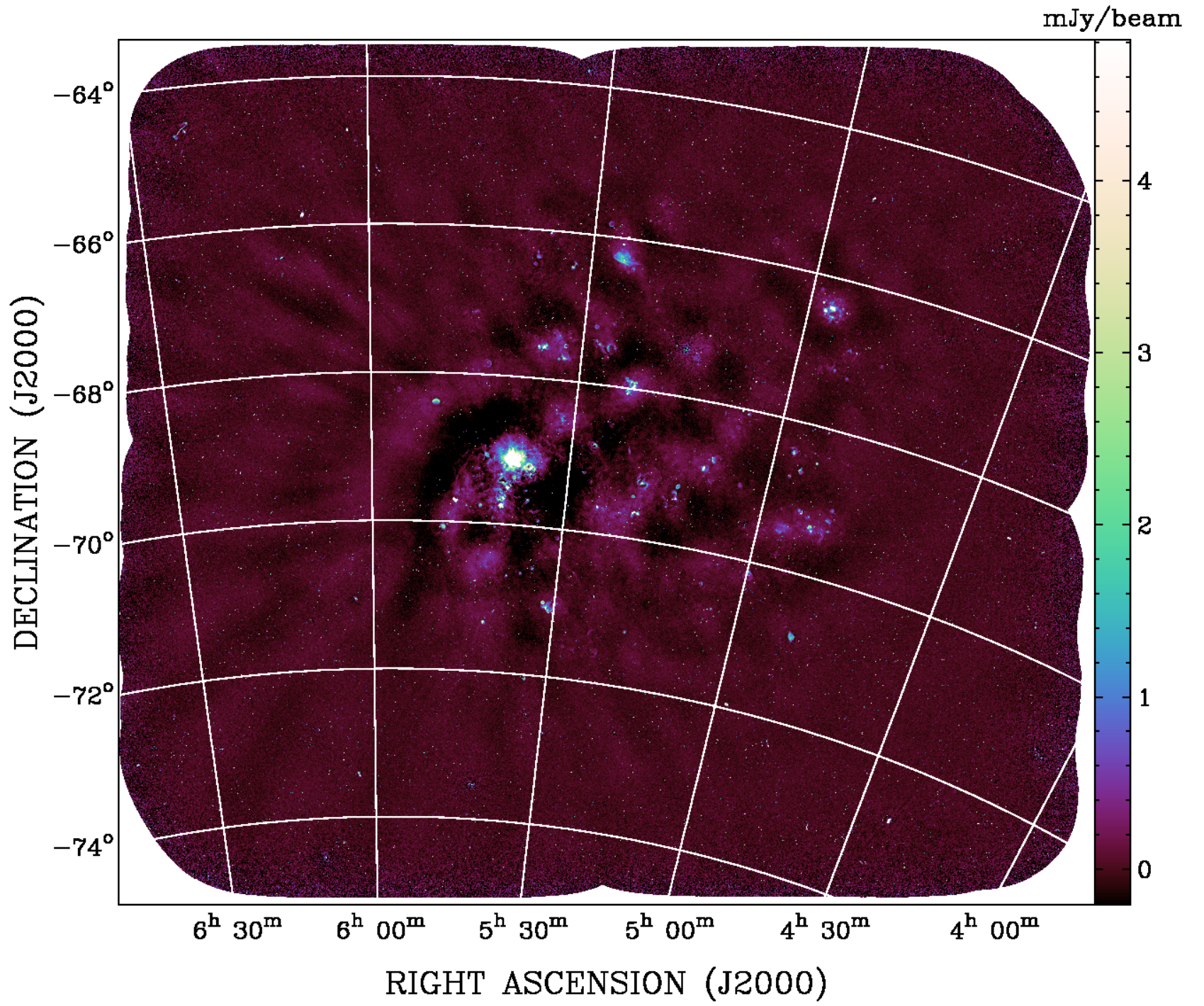}
	\end{tabular}
    \caption{ASKAP ESP image of the LMC at 888 MHz. The beam size is $13\rlap{.}^{\prime\prime}9\times12\rlap{.}^{\prime\prime}1$}
    \label{fig:LMCimage}
\end{figure}

\begin{figure}

	\includegraphics[width=\columnwidth]{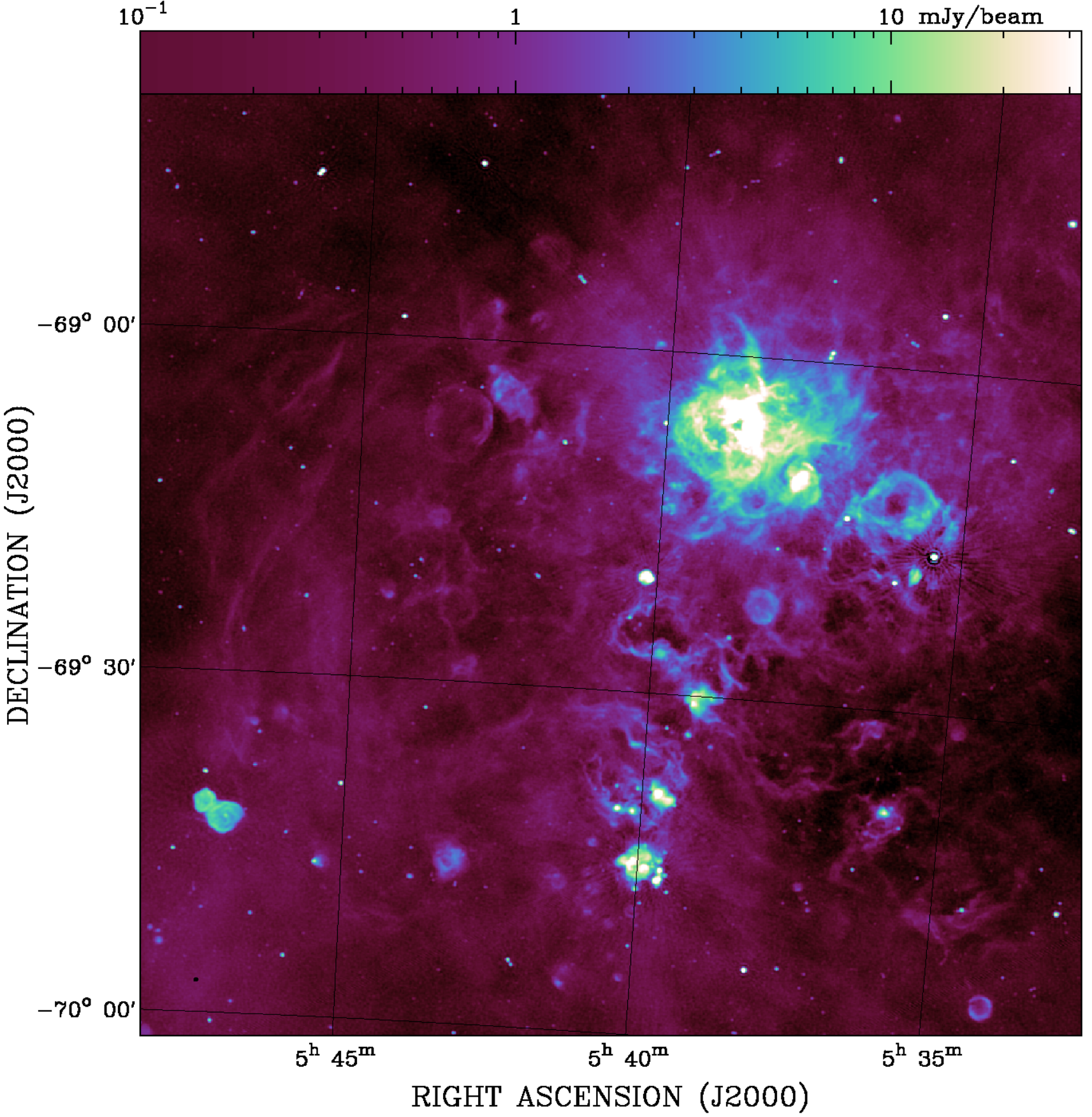}

    \caption{Region around 30\,Doradus from the ASKAP ESP image of the LMC.}
    \label{fig:LMC_30Dor}
\end{figure}

\begin{figure}

	\includegraphics[width=\columnwidth]{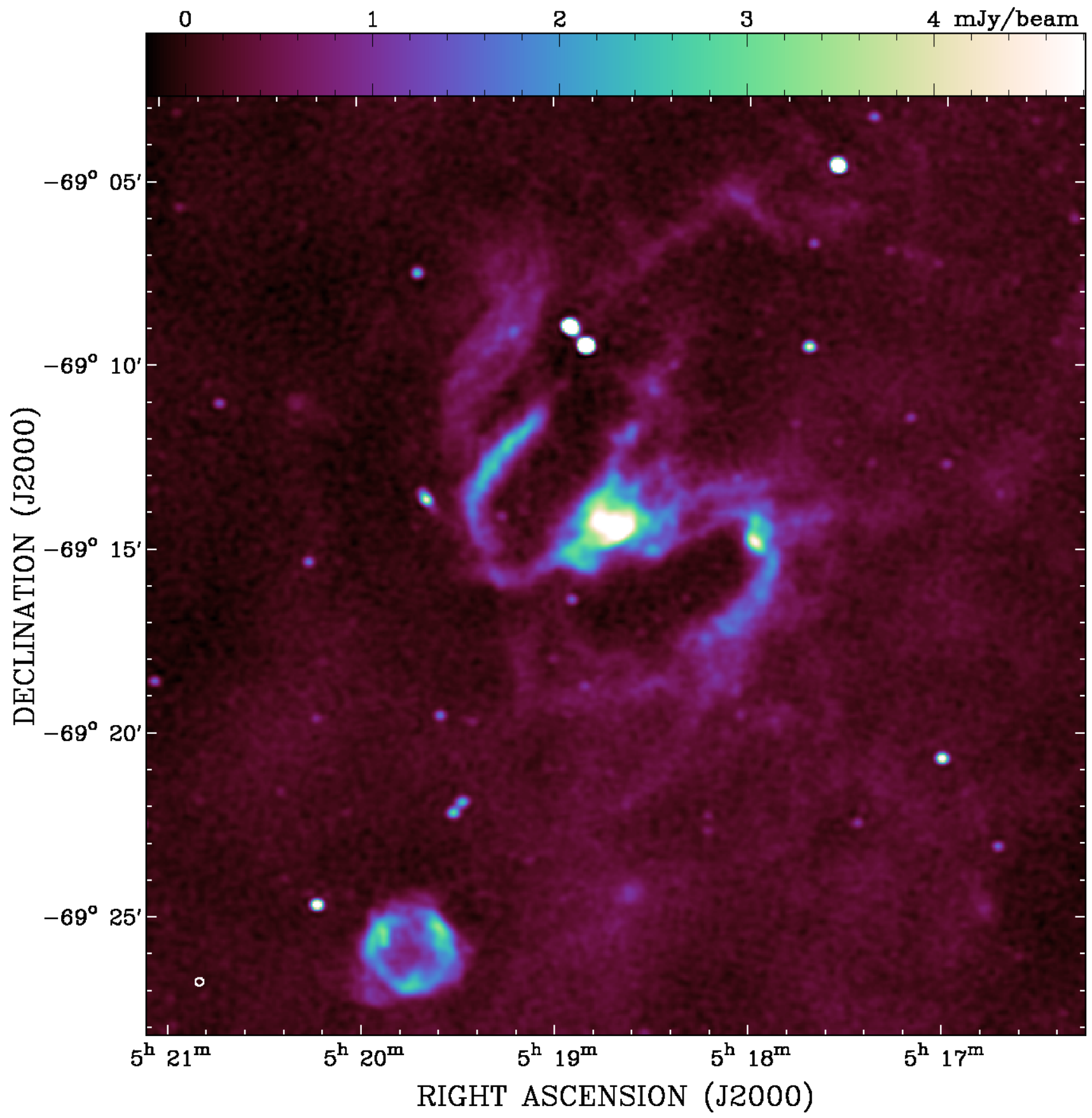}

    \caption{LHA\,120-N\,119 from the ASKAP ESP image of the LMC. N\,119 is an H\,{\sc ii} region close to the kinematic centre of the LMC, and has a pronounced spiral shape that is reminiscent of a barred spiral galaxy.}
    \label{fig:LMC_N119}
\end{figure}

Here we present an analysis of the observation of the LMC at 888 MHz with a bandwidth of 288 MHz taken on 2019 April 20 using ASKAP's full array of 36 antennas (scheduling block 8532). The LMC was observed as part of the ASKAP commissioning and early science (ACES, project code AS033) verification \citep{DeBoer2009,Hotan2014,McConnell2016} in order to investigate issues that were found in higher-frequency higher-spectral-resolution Galactic-ASKAP \citep[GASKAP;][]{Dickey2013} survey observations, as well as to test the rapid processing with ASKAPsoft \citep{Whiting2020}. The observations cover a total ﬁeld of view of 120 deg$^{2}$, with a total exposure time of $\sim$12h40m. It is compiled by four pointings ($\sim$3h10m each) with three interleaves\footnote{interleaves are overlapping pointings where the telescope slews between them at a more rapid cadence.} each to result in more uniform depth across the field -- effectively 12 pointings. The three interleaves overlap by $\sim$0.5 deg to improve the uniformity of sensitivity across the field. The ASKAP ESP image of the LMC can be seen in Figure \ref{fig:LMCimage}. Figures \ref{fig:LMC_30Dor} and \ref{fig:LMC_N119} show close ups of well known LMC regions 30\,Doradus (the Tarantula Nebula) and LHA\,120-N\,119, an H\,{\sc ii} region near the kinematic centre of the LMC. The largest angular scales that can be recovered in this survey are 25--50 arcmin \citep{McConnell2020}. The LMC exhibits structure on scales larger than can be recovered by the shortest baselines available with ASKAP (22 m). The missing short-spacing data results in negative bowls around some regions, for example, this is particularly evident around the extended bright structure in 30\,Doradus.
 
Although the image noise levels are near to expected values in most areas, a common feature of early ASKAP data has been the presence of some low-level artefacts (at $\sim$ 1 per cent peak brightness) very close to bright sources (a few hundred mJy and above). These artefacts generally appear as radial stripes and rapidly fade away from the source.
 
 Data processing was performed using ASKAPsoft \citep{Whiting2020} by the ASKAP operations team and the resulting images are available on the CSIRO ASKAP Science Data Archive \footnote{https://research.csiro.au/casda/}. Bandpass calibration was done using observations of PKS\,B1934$-$638, which establishes the flux scale and frequency-dependent complex gains for each of ASKAP's 36 beams. The bandpass-calibrated measurement set for each beam is imaged independently using a phase-only self-calibration approach.  The multi-scale, multi-frequency C\textsc{lean} algorithm was used to deconvolve the array response, with two Taylor terms representing spectral behaviour. Images for all beams were combined into a single field using a linear mosaic, correcting for the primary beam response.

The absolute flux density calibration uncertainty is tied to PKS B1934$-$638 (ATNF Technical Document Ser. 39.3/040\footnote{ https://www.atnf.csiro.au/observers/memos/d96783~1.pdf}), which differs by no more than 1--2 per cent compared to the Baars scale \citep{Baars1977}. Additional uncertainties are introduced by point spread function (PSF) variation in a mosaic, generally $\sim$ 1--2 per cent for long-track observations such as this one. In later observations this is alleviated by convolving each beam image to the smallest possible common PSF \citep{McConnell2020}. Uncertainties are also introduced by the assumption that all beams have a consistent Gaussian primary beam-shape. In reality holography has shown that they are somewhat non-Gaussian (particularly edge and corner beams of the phased-array instantaneous field) differing from a Gaussian shape by up to 8 per cent \citep{McConnell2020}. Therefore, the overall absolute flux density calibration uncertainty is 8 per cent.

\section{ASKAP ESP LMC Source List}
\subsection{Source detection}
We created the 888 MHz LMC ASKAP source catalog using the A\textsc{egean} source finder algorithm \citep{Hancock2012,Hancock2018}. Variable noise is present across the image, from the combination of the multiple beams and artefacts from bright sources. To ensure accurate source thresholds the variable noise must be parameterised before source finding. The B\textsc{ane} (Background and Noise Estimation) routine in the AegeanTools software was used, with default parameters, to create root mean squared noise (RMS) and background level maps\footnote{It does this via a sliding boxcar approach where it calculates the background and noise properties of all pixels within a box centred on a given grid point, the resulting maps having the same pixel scale as the input images.}. The maps were then used, alongside the LMC image, with A\textsc{egean}, using default parameters to create initial source lists. A cursory visual inspection of the sources was carried out to verify detections from the initial source list.

\begin{figure}

	\includegraphics[width=\columnwidth]{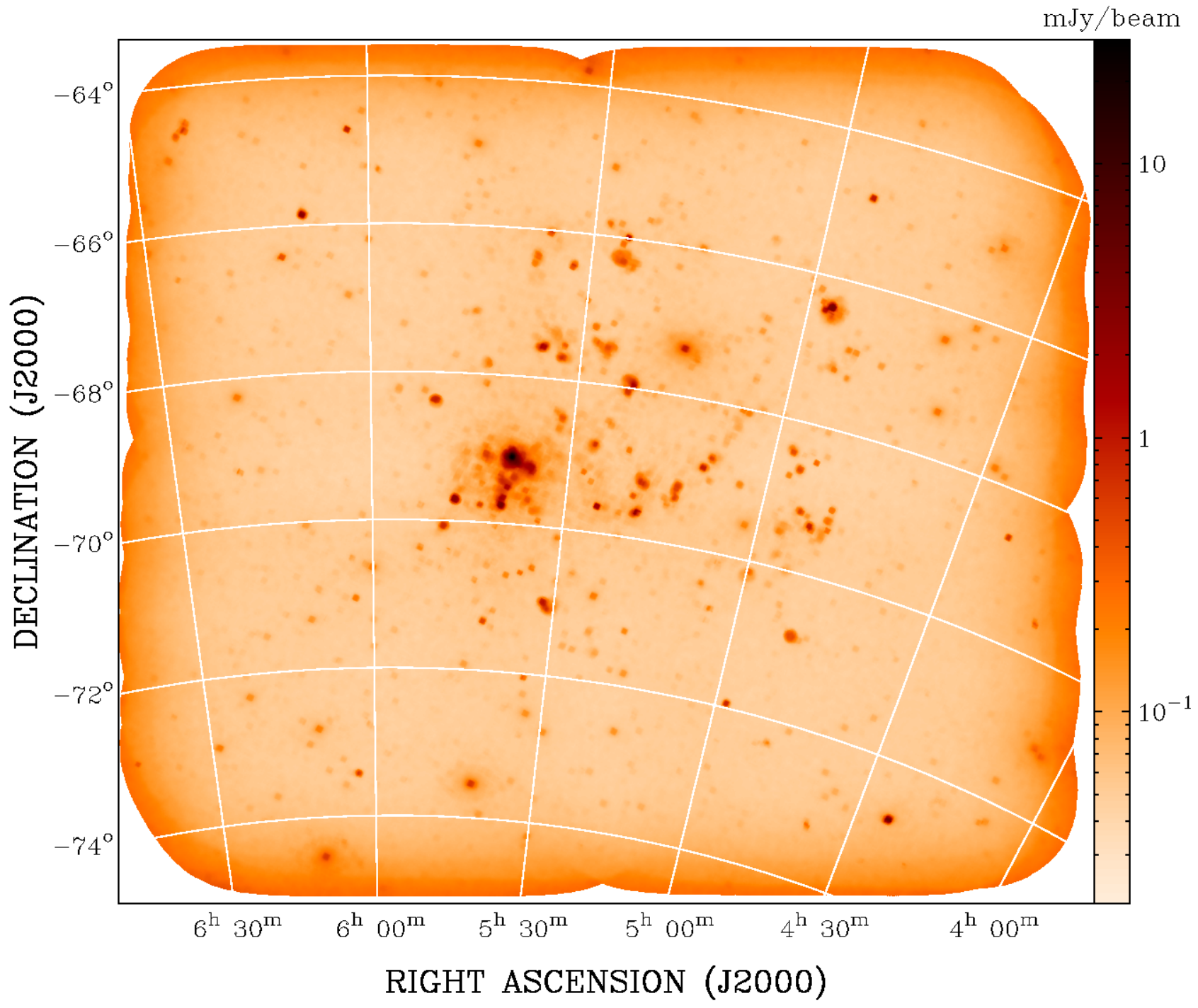}

    \caption{RMS map of the 888 MHz ASKAP observations, produced by B\textsc{ane} with the default parameters. The image is of the same pixel scale as in Figure \ref{fig:LMCimage}. Higher RMS levels are found at the edge of the field (due to fewer beams present and because of the primary beam shape of which the sensitivity falls off) and around the brighter sources. }
    \label{fig:RMSmap}
\end{figure}

To ascertain the best  A\textsc{egean} parameters, the algorithm was used on subsamples of the image with different values of \textsc{seedclip} (the number of $\sigma$ above the local RMS for sources to be deemed detected) to find the best \textsc{seedclip} value that obtains the least false positive detections without sacrificing too many true positive detections. To quantify the false detection rate we can assume that the noise in the image follows a symmetric, Gaussian distribution. This symmetry means that the negative image (inverted) has the same noise properties as the non-inverted image, meaning large noise troughs in the original image are now detected as sources. Hence, the number of detected sources should be approximately equal to the number arising from the false positive noise spikes in the true image.
From this we can constrain the “real” detections in an image by investigating how the false detections vary with \textsc{seedclip} and then pick out the optimum value that successfully extracts sources with minimal contamination from noise \citep{Hale2019}. The percentage of real detections can therefore be calculated using
\begin{equation}
\%Real Detections = 100 \times \frac{N_{+}-N_{-}}{N_{+}}
\end{equation}
where $N_{+}$ and $N_{-}$ are the total number of sources detected with A\textsc{egean} on the original and inverted images, respectively. Testing proved that for a large variety of regions, containing from compact to large extended sources, the default value of $5\sigma$ was appropriate.

\subsection{Source lists}
In total, we found 54,612 sources in the ASKAP 888 MHz image. We separated this source list into a G\textsc{old} (30,866), S\textsc{ilver} (22,080) and B\textsc{ronze} (1,666) source list. Some sample records of the G\textsc{old} source list can be found in Table \ref{ex_table}. The PSF was generated during image restoration, which means that the semi-major axis ($13\rlap{.}^{\prime\prime}9$), semi-minor axis ($12\rlap{.}^{\prime\prime}1$) and position angle ($-84\rlap{.}^{\circ}4$) of the PSF are constant for all sources. The sizes of the sources were determined during source finding. All sources have a designated 'Island' number, and if multiple sources have the same 'Island' number then they are considered components of the island and will have different 'Source' numbers. These components of the islands can be spread across the different source lists. The isolated (one component in an island) sources make up $\sim$ 90 per cent of the entire catalogue. We note that isolated sources have more reliable flux density errors than sources that are components of islands, which tend to have unreasonably small flux density errors calculated by A\textsc{egean}.

\begin{table*}
\caption{Example of the catalogue created from the LMC ASKAP ESP image. In the second part of the "EMU ID" the sources are labelled either as "ES" or "EC", where "E" stands for early science data, "S" stands for an isolated source and "C" stands for a component of a source. The $\sigma$ indicates the source-finding fitting error. "Peak" indicates the peak flux density per beam and error. "Integrated" indicates the integrated flux density, which is not fit directly but is calculated from {\it a}, {\it b} and peak flux and the synthesised beam size. The {\it a} is the fitted semi-major axis, {\it b} is the fitted semi-minor axis and PA is the fitted position angle. "Island" is the number of the island from which the source was fit and the "Source" is the number within that island, starting from zero for components of a source (EC) and always zero for isolated sources (ES). "Background" is the flux density per beam measured from the background map created with B\textsc{ane} at the source position. "Local RMS" is the flux density per beam measured from the rms map created with B\textsc{ane} at the source position. "Residual Flux" mean and standard deviation are the mean and standard deviation of the residual flux remaining in the island after fitted Gaussian is subtracted.  Flags are 0 for a good fit, 2 indicates an error occurred during the fitting process (e.g., the fit did not converge), 4 indicates a component was forced to have the shape of the local point spread function. }
\begin{tabular}{lcccccccccccccccc}
\hline\hline
Source List & EMU ID & RA (J2000) & Dec (J2000) & \multicolumn{2}{c}{----- peak -----} & \multicolumn{2}{c}{--- integrated ---} & \\
 & & $^h$ $^m$ $^s$ & $^\circ$ $^\prime$ $^{\prime\prime}$ & $F$ (mJy beam$^{-1}$) & $\sigma$ (mJy beam$^{-1}$) & $F$ (mJy) & $\sigma$ (mJy)  \\
\hline
G\textsc{old} & EMU ES J063120.8$-$741322 & 06:31:20.8 & $-$74:13:23 & 6.20 & 0.11 & \mbox{~~}7.60 & 0.14 & \mbox{~~}\mbox{~~}\mbox{~~}\mbox{~~}\mbox{~~}\mbox{~~}\\
G\textsc{old} & EMU ES J063111.1$-$741304 & 06:31:11.3 & $-$74:13:04 & 1.65 & 0.08 & \mbox{~~}4.56 & 0.25 & \\
G\textsc{old} & EMU ES J062722.6$-$742301 & 06:27:22.6 & $-$74:23:01 & 1.46 & 0.11 & \mbox{~~}1.67 & 0.14 & \\
G\textsc{old} & EMU ES J062920.3$-$741744 & 06:29:20.4 & $-$74:17:45 & 0.74 & 0.11 & \mbox{~~}0.75 & 0.12 & \\
G\textsc{old} & EMU ES J063221.4$-$740905 & 06:32:21.4 & $-$74:09:06 & 9.74 & 0.11 & 11.40 & 0.14 & \\
... \\
\hline
\end{tabular}
\label{ex_table}
\end{table*}

\begin{table*}
\contcaption{}
\begin{tabular}{cccccccccccccc}
\hline\hline
\multicolumn{2}{c}{----- a -----} &  \multicolumn{2}{c}{----- b -----} & \multicolumn{2}{c}{----- PA -----} & Island & Source & Background & Local RMS & \multicolumn{2}{c}{Residual Flux} & flags & \mbox{~~}\mbox{~~} \\
($^{\prime\prime}$) & $\sigma$ ($^{\prime\prime}$) & ($^{\prime\prime}$) & $\sigma$ ($^{\prime\prime}$) & $^\circ$ & $\sigma$ ($^\circ$) & & & (mJy beam$^{-1}$) & (mJy beam$^{-1}$) & mean (mJy) & $\sigma$ (mJy) & & \\
\hline
15.49 & 0.07 & 13.51 & 0.06 & $-$87.07 & 0.10 & \mbox{~~}\mbox{~~}113 & 0 & 0.009 & 0.114 & $-$0.038 & 0.146 & 0 &\\
27.27 & 0.39 & 17.28 & 0.23 & $-$87.08 & 0.08 & \mbox{~~}\mbox{~~}115 & 0 & 0.007 & 0.111 & \mbox{~~}0.007 & 0.189 & 0 &\\
16.18 & 0.32 & 12.05 & 0.22 & $-$86.66 & 0.19 & \mbox{~~}\mbox{~~}117 & 0 & 0.015 & 0.118 & $-$0.013 & 0.075 & 0 &\\
14.12 & 0.65 & 12.21 & 0.56 & $-$78.88 & 0.82 & \mbox{~~}\mbox{~~}118 & 0 & 0.007 & 0.113 & $-$0.006 & 0.025 & 0 &\\
15.10 & 0.04 & 13.24 & 0.04 & $-$84.38 & 0.07 & \mbox{~~}\mbox{~~}119 & 0 & 0.004 & 0.116 & $-$0.010 & 0.100 & 0 &\\
... \\
\hline
\end{tabular}
\end{table*}

To remove multiple detections of large/bright extended sources, a threshold of the local RMS was used. This is because in the area of these large sources the value of local RMS is higher than elsewhere. A cut of RMS $<$ 0.12 mJy was used and a cursory visual inspection was undertaken of the sources removed (and not removed) to verify the effectiveness of this cut. This cut also had the effect of removing sources close to/surrounded by bright extended sources, as well as the sources along the outermost edges of the image due to the poorer sensitivity in this part of the image. 

\begin{figure*}
\centering
    \begin{tabular}{c}
	\includegraphics[trim={0 1cm 0 1cm},width=\textwidth]{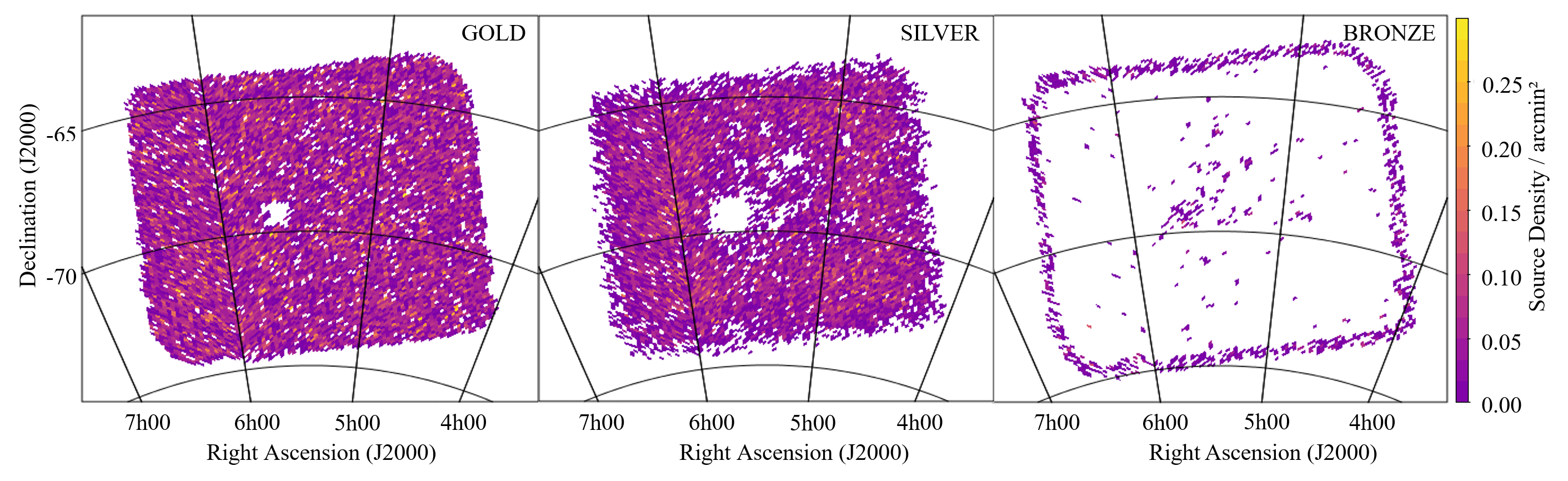}
	\end{tabular}
    \caption{Density plots of the G\textsc{old} (left), S\textsc{ilver} (centre) and B\textsc{ronze} source lists (right).  This shows that the G\textsc{old} source list is uniform across the LMC field, except in the area of 30\,Doradus, which is seen as a blank space. The S\textsc{ilver} source list is not as uniform and showcases gaps where the brighter Magellanic sources are present.  The B\textsc{ronze} source list covers the edges of the LMC field and the areas of the brightest sources, where the local RMS is greatest.}
    \label{fig:densityplots}
\end{figure*}

\begin{figure}
	\includegraphics[trim={0.5cm 1.5cm 0.5cm 1cm},width=\columnwidth]{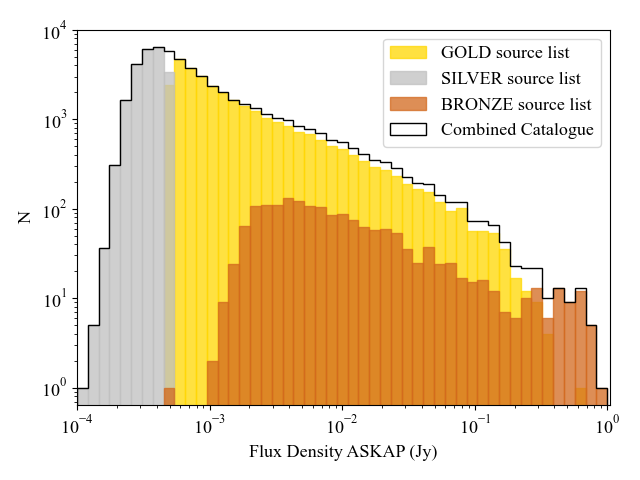}
    \caption{Histogram of integrated flux density at 888 MHz illustrating the G\textsc{old}, S\textsc{ilver}, and B\textsc{ronze} populations.}
    \label{fig:flux_hist}
\end{figure}

The G\textsc{old} standard source list was selected to be a spatially uniform distribution of sources, so that the fainter sources, that are not uniform in spatial distribution, are separated into a S\textsc{ilver} source list (see Figure \ref{fig:densityplots} (left \& centre)). For a source to be moved to the S\textsc{ilver} list an integrated flux density of at most 0.5 mJy was required. This value was chosen because this is where the flux density distribution of the ASKAP radio sources turns over, see Figure \ref{fig:flux_hist}. The B\textsc{ronze} source list was created, which consists of the sources that were removed with the local RMS cutoff but judged subsequently to be true sources. Each source was individually visually inspected using SAOImageDS9 \citep{Joye2003}, where a source was kept if it was clearly visible above the background noise and was point-like or slightly resolved. This reclaimed some of the sources near the edge of the image where the high noise removed them from the S\textsc{ilver} and G\textsc{old} lists, as well as the sources that were near bright extended emissions, including well-known bright sources such as the supernova SN\,1987A. The sources that were not recovered are those located in complex regions or blended with other sources that would require higher angular resolution to disentangle. The flux densities of a few sources in noisy or confused regions may be unreliable. Therefore, it is advised that for these sources the user inspects the image and ascertain whether to trust the given flux density values. Of the 4,072 sources originally removed, 1,666 were placed in the B\textsc{ronze} source list, the density map of which is presented in Figure \ref{fig:densityplots} (right).

\subsection{Comparison with previous radio catalogues}

The ASKAP ESP source lists were compared with various radio continuum catalogues that cover the LMC. The number counts for our survey were compared with those from the previous ASKAP ESP catalogue of the SMC \citep{Joseph2019}, for which the full ASKAP array was not available. See Table \ref{table:Radio_surveys_comp} for a comparison of previous radio surveys of the LMC.

\begin{table*}
\caption{Table comparing existing radio surveys of the LMC with the survey presented here.}
\begin{tabular}{ l c c c c c c c }
        \hline\hline
        Survey & Telescope & Frequency & RMS & Beam Size & Bandwidth & Source Count & Reference  \\
         &  &  GHz & mJy beam$^{-1}$ & $^{\prime\prime}$ & MHz & &\\
        \hline
        EMU ESP & ASKAP & 0.888 & 0.0577 & 13.9 $\times$ 12.1 & 256 & 53547 & This Work\\
        ASKAP-Beta & ASKAP & 0.843 & 0.71 & 61 $\times$ 53 & 1 & 1995 & \cite{FilipovicTest} \\
        1.384 GHz ATCA & ATCA & 1.384 & 0.5 & 40 & 128 & 6623 & \cite{Hughes2007} \\
        4.8 / 8.64 GHz ATCA & ATCA & 4.8 / 8.64 & 0.28 / 0.5 & 33 / 20 &  128 & 801 / 419 & \cite{Dickel2005} \\
        SUMSS & MOST & 0.843 & $\sim$ 1 & 43 & 3 & 3829 & \cite{Mauch2003} \\
        PMN Survey & Parkes & 4.85 & $\sim$ 4.2 & 126 & 600 & 410 & \cite{Griffith1993} \\
        843 MHz MOST & MOST & 0.843 & 0.3 -- 0.4 & 43 & 3 & 2162 & \cite{Mills1985} \\
        \hline
        \label{table:Radio_surveys_comp}
\end{tabular}
\end{table*}


\subsubsection{SUMSS 843 MHz}
The MOST telescope was used to carry out the Sydney University Molonglo Sky Survey \citep[SUMSS;][]{Mauch2003, Mauch2003cat} at 843 MHz with a beam size of $45^{\prime\prime}\times45^{\prime\prime}$ to create a catalogue that covers the whole sky South of DEC $= -$30~deg, including the LMC. The positions in the catalogue are accurate to within 1--2~arcsec for the brighter sources ($>$ 20 mJy beam$^{-1}$) and always better than 10~arcsec. This catalogue therefore makes for a suitable comparison for the new 888 MHz ASKAP catalogue.

\begin{figure}

	\includegraphics[trim={0.5cm 1cm 0.5cm 0.5cm},width=\columnwidth]{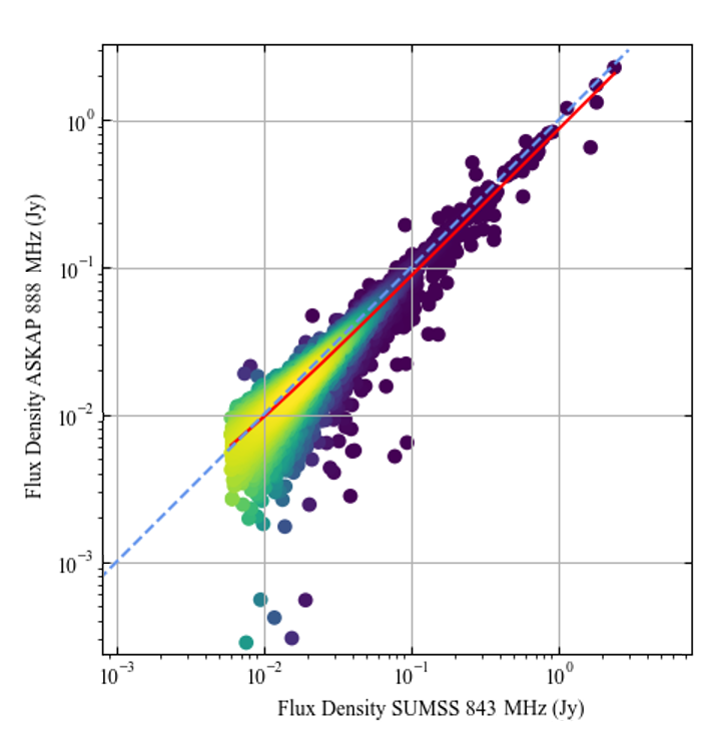}
    \caption{Comparison between the integrated flux densities calculated for the ASKAP 888 MHz image, and those from calibrated archival data of SUMSS MOST 843 MHz data. The blue dashed line indicates the one-to-one relation and the red line is the line of best fit. This shows a tight correlation. The outliers in the lower part of the figure are the single SUMSS sources that were resolved by ASKAP into multiple components.}
    \label{fig:flux_comp}
\end{figure}

The new ASKAP 888 MHz source lists (G\textsc{old} and S\textsc{ilver}, 53,547 sources) were cross-matched with the SUMSS 843 MHz catalogue (3,829 sources), with a search radius of 10 arcsec, yielding 3,211 matches. The flux densities are compared in Figure \ref{fig:flux_comp}; the dashed blue line indicates the one-to-one relation and the red line indicates the line of best fit. The sources that show higher than expected flux density from ASKAP compared to SUMSS were found to be near bright point sources with nearby artefacts which, when coinciding with a fainter source, increases their total flux density. The sources that show lower than expected flux density from ASKAP compared to SUMSS were found to be extended sources, that were seen as one source in SUMSS but multiple components in ASKAP, hence the lower flux density. The correlation between the calculated integrated flux densities of ASKAP 888 MHz and SUMSS 843 MHz source lists is described by the Pearson Correlation Coefficient, $r=0.967 \pm 0.004$, showing a strong positive correlation and that LMC ASKAP is well flux calibrated. 

In Figure \ref{fig:flux_comp} the red line is slightly "above" the density ridge of the source distribution, so could be interpreted by the expectation that the small frequency difference from 843 to 888 MHz causes on average a lower flux density at 888 MHz. For $\alpha= -0.7$  (typical of synchrotron emission) this would be $(888 MHz/843 MHz)^{-0.7} = 0.96$, whereas the true mode of the integrated flux density ratio is ($0.946 \pm 0.005$), which is only a little smaller than expected.

\begin{figure}

	\includegraphics[trim={0 1.5cm 0 1cm},width=\columnwidth]{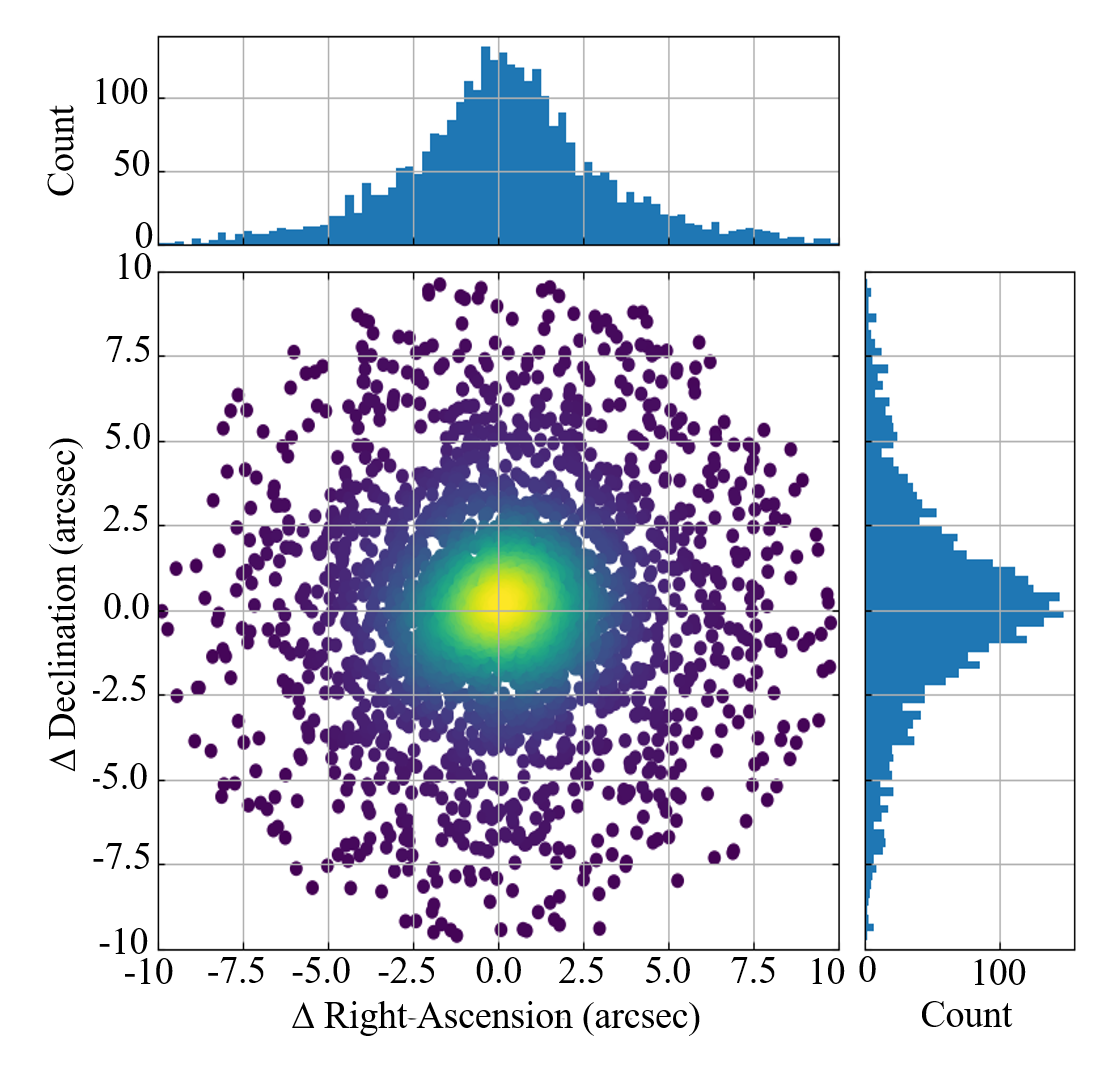}
    \caption{Figure of positional difference (MOST -- ASKAP) of the 2,949 radio sources from the ASKAP (888~MHz) catalogue and the SUMSS (843~MHz) catalogue.}
    \label{fig:pos_comp}
\end{figure}

We compare the positional differences ($\Delta$RA and $\Delta$DEC) between our new ASKAP catalogue and the previous SUMSS catalogue at 843 MHz in Figure \ref{fig:pos_comp}. This figure shows that the positional offsets are well centred, with the mean offsets calculated to be $(0.189 \pm 0.054)^{\prime\prime}$ and $(0.088 \pm 0.054)^{\prime\prime}$ for RA and DEC, respectively. Overall, the ASKAP LMC astrometry is well-calibrated.

\subsubsection{ASKAP SMC ESP}

Two radio continuum images from the ASKAP survey in the direction of the SMC were taken as part of the EMU Early Science Project (ESP) survey of the Magellanic Clouds \citep{Joseph2019}. The 888 MHz observation of the LMC covers four times the area of the ASKAP SMC ESP observations. The two source lists that were produced from these images by \cite{Joseph2019} contain radio continuum sources observed at 960 MHz (4489 sources) and 1320 MHz (5954 sources) with a bandwidth of 192 MHz and beam sizes of $30^{\prime\prime}\times30^{\prime\prime}$ and $16\rlap{.}^{\prime\prime}3\times15\rlap{.}^{\prime\prime}1$, respectively. The median RMS noise values were 186 $\mu$Jy beam$^{-1}$ (960 MHz) and 165 $\mu$Jy beam$^{-1}$ (1320 MHz). The observations of the SMC were made with only 33 per cent and 44 per cent (for 960 MHz and 1320 MHz respectively) of the full ASKAP antenna configuration and 66 per cent of the final bandwidth that was available in the final array, with which the LMC was observed.

\begin{figure}

	\includegraphics[trim={0 1cm 0 1cm},width=\columnwidth]{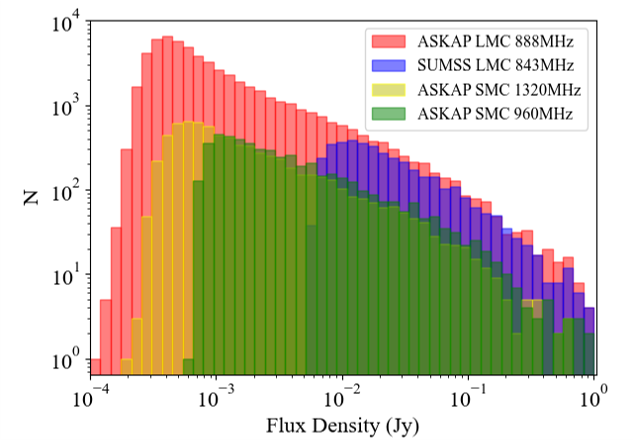}
    \caption{Integrated flux density histograms of ASKAP LMC (888 MHz), ASKAP SMC (960 MHz \& 1320 MHz) and SUMSS LMC (843 MHz). The new ASKAP LMC image shows a marked improvement in sensitivity for a radio survey of the LMC and compared to the ASKAP SMC survey.}
    \label{fig:flux_cat_comp}
\end{figure}

Figure \ref{fig:flux_cat_comp} shows the number of sources as a function of flux density for ASKAP LMC (888 MHz), ASKAP SMC (960 MHz and 1320 MHz) and SUMSS LMC (843 MHz). As expected, the number of sources is lower and the detection limit is higher in the ASKAP SMC catalogue compared to the ASKAP LMC catalogue. This is due to the smaller survey area, the differences in sensitivity for each image and the observing frequency.

\begin{figure}

	\includegraphics[trim={0 1.5cm 0 1cm}, width=\columnwidth]{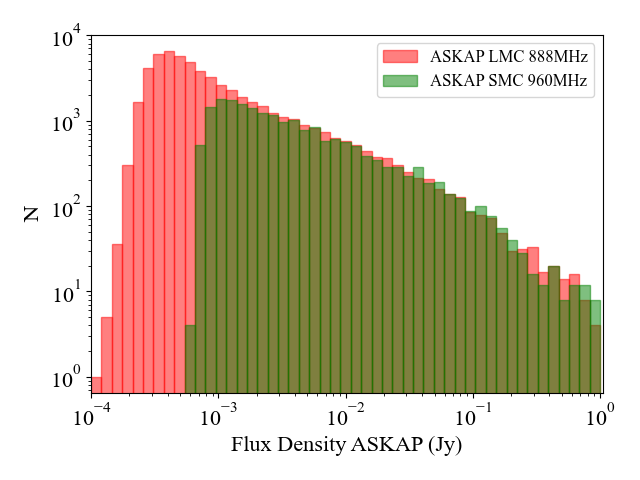}
    \caption{Comparison between the integrated flux densities calculated for the ASKAP images, of the LMC and SMC at 888 MHz and 960 MHz, respectively. The SMC numbers have been scaled to the same of the LMC image for better comparison.}
    \label{fig:flux_comp_SMCLMC}
\end{figure}

 The comparison of the SMC (960 MHz) sources with the LMC sources (888 MHz) where the SMC is scaled up to the same area size as the LMC (see Figure \ref{fig:flux_comp_SMCLMC}), shows increased sensitivity with the array fully operational and that within the flux limits of the SMC both LMC and SMC ASKAP fields show a similar source density. In contrast to
the SMC observation, the LMC observation goes deep enough to surpass the
upturn in the luminosity function and probes the radio-faint population
(for more details see section \ref{Backgroundsources}). Overall, this shows the improvement of using the fully operational ASKAP array. Lastly, the LMC and SMC (the foreground sources) do not fill the area (at least not at high source density) of the ASKAP images, which means that the Clouds do not contribute a significant fraction to the predominantly extragalactic source density.

\section{Magellanic and Galactic sources}

We discuss a few types of Magellanic and Galactic sources here. These sources were found by cross-matching with various catalogues using T\textsc{OPCAT}\footnote{http://www.star.bris.ac.uk/$\sim$mbt/topcat/} \citep{Taylor2005} with an initial search radius of 10 arcsec\footnote{to allow for inaccuracies in both the ASKAP and literature positions.}, after which visual inspection was used to determine whether they are true counterparts. All quoted flux densities from the comparison catalogues are integrated flux densities unless stated otherwise.


\subsection{Planetary Nebul{\ae} and other emission-line objects}

\cite{Leverenz2017} \citep[see also][]{Filipovic2009} presented radio
detections of 28 PNe in the LMC, but these were mostly detected at 4.8 GHz
(26) with only 14 detected at 1.4 GHz as the radio emission is dominated
by free--free emission. Simbad lists 1,334 objects within the ASKAP footprint that have at one time been classed as a PN or candidate PN. Of these, 114
have counterparts in our ASKAP catalogue within 10 arcsec. An example of this list of objects is shown in Table \ref{PNetable} (the full table is available online as a supplementary material). Note that in some cases we believe the (candidate) PN is not the radio counterpart, especially for positional differences of $>$ 6 arcsec.

\begin{table*}
\caption{Example of ASKAP 888 MHz radio detections within 10 arcsec of sources at one time classed as potential PNe. The positions are those of the radio source, and the distance ($\Delta$) is that between the radio source and the coordinates listed in Simbad. The most likely classes are Planetary Nebula (PN), candidate post-AGB object (pAGB), emission-line star (em$*$), Young Stellar Object (YSO) or H\,{\sc ii} region (H\,{\sc ii}). See comments where there is doubt. The full table is available online in the supplementary material.}
\begin{tabular}{lccccccccll}
\hline\hline
Name & EMU ID & RA (J2000) & Dec (J2000) & \multicolumn{2}{c}{----- peak -----} & \multicolumn{2}{c}{--- integrated ---} & $\Delta$ & Class & Comment\rlap{s} \\
 & & $^h$ $^m$ $^s$ & $^\circ$ $^\prime$ $^{\prime\prime}$ & \llap{$F$} (mJy) & $\sigma$ (mJy\rlap{)} & \llap{$F$} (mJy) & $\sigma$ (mJy\rlap{)} & $^{\prime\prime}$ & & \\
\hline
SMP\,LMC\,10\rlap{4}      & EMU ES J042437.3$-$694221  & 04 24 37.4 & $-$69 42 21 &         0.31 & 0.06 &         0.34         & 0.06         & 0.9 & PN?         & \\
SMP\,LMC\,6        & EMU ES J044738.8$-$722821 & 04 47 38.8 & $-$72 28 21 &         0.80 & 0.05 &         0.78         & 0.05         & 0.9 & PN          & \\
SMP\,LMC\,5        & EMU ES J044808.4$-$672606 & 04 48 08.4 & $-$67 26 07 &         0.39 & 0.05 &         0.34         & 0.04         & 0.7 & pAGB        & \\
SMP\,LMC\,7        & EMU ES J044829.6$-$690813 & 04 48 29.6 & $-$69 08 13 &         0.45 & 0.05 &         0.48         & 0.05         & 0.5 & PN          & \\
SMP\,LMC\,10       & EMU ES J045108.7$-$684904 & 04 51 08.7 & $-$68 49 05 &         0.45 & 0.05 &         0.47         & 0.05         & 1.3 & PN          & \\
Sk\,$-$70\,10      & EMU EC J045325.9$-$703541 & 04 53 25.9 & $-$70 35 42 &         0.44 & 0.09 &         1.45         & 0.30         & 3.1 & H\,{\sc ii} & Contains  \\
& & & & & & & & & & B0\,III[e] \\
... & & & & & & & & & & \\
\hline
\end{tabular}
\label{PNetable}
\end{table*}

However, only 43 of these have a primary Simbad class of PN (or candidate
PN) and two of these turn out not to be PNe (extended radio emission) and another four
have highly uncertain radio detections (a noise spike in one case, two cases of blends, and HD\,269404 described below), leaving 37 good radio detections
of likely PNe. Another 23 are classed as candidate post-asymptotic giant branch (post-AGB) objects, of which one has an
uncertain radio detection (and is discarded) and another one is classed as a carbon-rich
Wolf--Rayet star.

This radio sample of 114 radio sources also includes 31 emission-line stars and (three) YSOs, of which two are discarded because of their large positional differences $> $9 arcsec, and another 19 compact H\,{\sc ii} regions or other extended and/or bright radio sources that are likely also H\,{\sc ii} regions. It is unclear
whether there are any post-AGB objects among the emission-line stars.

Among the more interesting sources are HD\,269404 =
IRAS\,05216$-$6753 ($F_{888}=3.77\pm0.11$ mJy) -- still listed in Simbad
as a PN but probably a young, massive dust-enshrouded star \citep{vanLoon2001,vanLoon2010}; and LMC\,SySt-23 ($F_{888}=0.37\pm0.05$ mJy) -- a confirmed
D-type symbiotic binary \citep{vanAarle2011,Akras2019}. One
of the most interesting true PNe is SMP\,LMC\,83 ($F_{888}=1.36\pm0.07$
mJy) -- with a nitrogen-rich Wolf--Rayet type central star ([WN4.5:])
betraying a massive AGB star progenitor commensurate with it being the
radio loudest PN in this sample.

\begin{figure}

	\includegraphics[trim={0.5cm 5.5cm 0.5cm 5.5cm},width=\columnwidth]{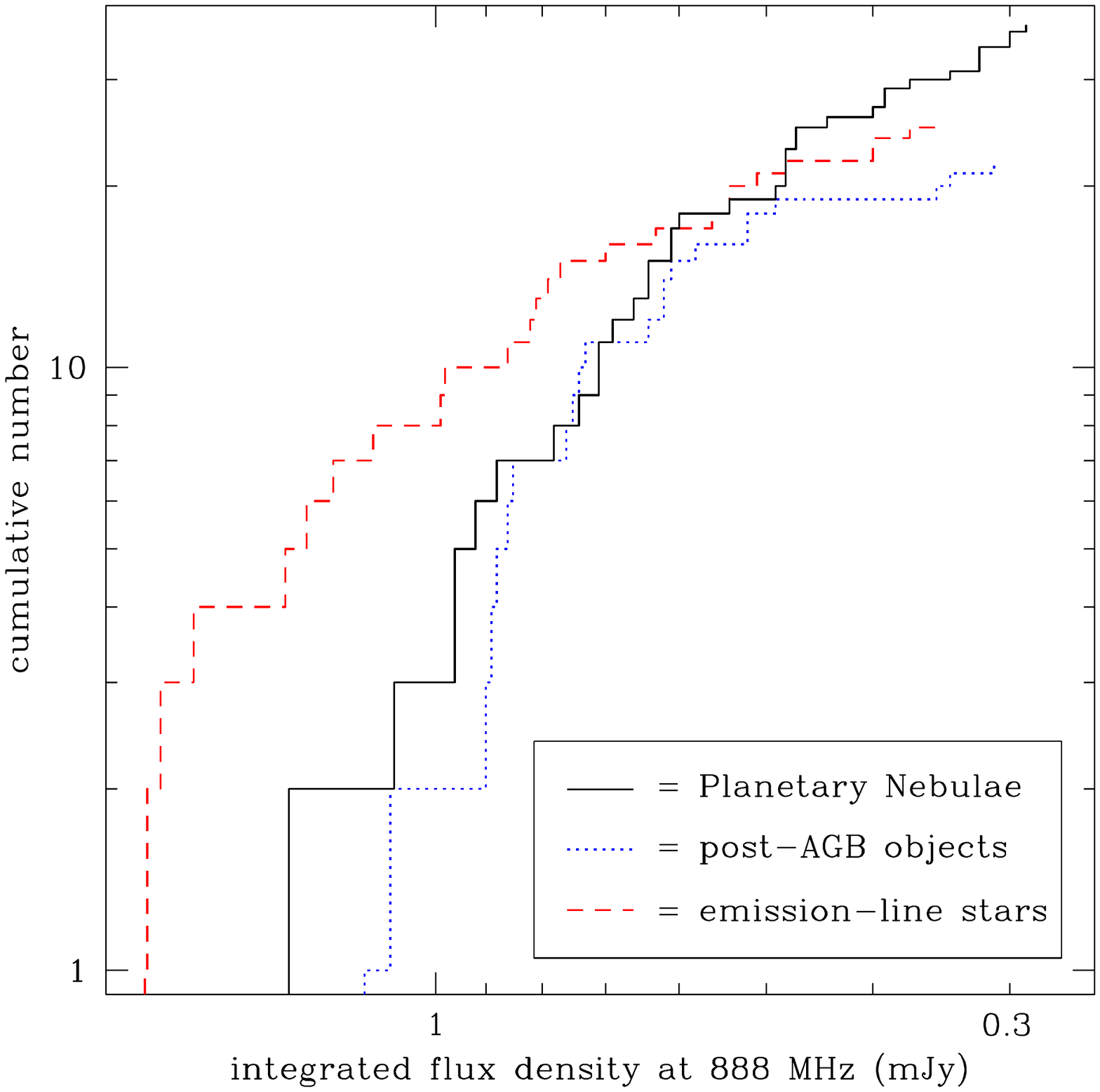}

    \caption{Cumulative 888 MHz flux distributions of planetary nebul\ae, post-AGB objects and emission-line stars.}
    \label{fig:PNe_cum_rad_lum_func}
\end{figure}

The cumulative histogram of integrated flux densities of the PNe (Figure \ref{fig:PNe_cum_rad_lum_func}) rises rapidly towards the faint end. The turnover of the 
distribution at
$F_{888}\sim0.5$ mJy can be attributed to incompleteness; a similar turnover can be
seen in the integrated flux density distribution of candidate post-AGB objects. However, the candidate
post-AGB distribution is less steep at the faint side of the turnover,
compared to the PNe. This may reflect the evolutionary
timescale, with a low-mass post-AGB system evolving more slowly and
therefore not ionizing its circumstellar envelope as efficiently. Possibly, the brighter sources descend from massive
AGB progenitors and are surrounded by a larger ionized envelope and/or a
higher electron density compared to lower-mass systems. The distinction between post-AGB and PN is not sharply defined, though, and some
of the post-AGB objects may well be considered to be PNe.

The emission-line star integrated flux densities function (Figure \ref{fig:PNe_cum_rad_lum_func}) resembles that of the candidate post-AGB stars at the faint end. At the bright end, however, there is a clear excess of emission-line stars
over PNe/post-AGB candidates. Given that the H\,{\sc ii} sources within
this compilation of 114 radio sources also often had emission-line stars associated with them,
it is likely that those bright emission-line stars are massive stars. It
is not clear whether these are young or evolved, but one could speculate
that young sources are more often associated with extended radio emission
from an H\,{\sc ii} region and evolved stars are more often isolated and
thus unresolved. This is somewhat corroborated by the fact that of all
stars with supergiant luminosity class (I or II, according to Simbad), five are unresolved by
ASKAP and only one is found in extended radio emission, versus a ratio of
20:11 for luminosity classes III--V. Forbidden line emission ("[e]") is
marginally more prevalent among unresolved radio sources (15 out of 27) as
well, compared to stars associated with extended radio emission (5 out of
11).


\subsection{Young stellar objects}

YSOs become radio continuum sources once they
ionize a sufficient amount of gas and free--free emission yields a
relatively flat continuum emission spectrum extending from the near-IR far
into the radio regime. They are also found in complex star-forming regions
that are often embedded in H\,{\sc ii} regions that exhibit bright free--free radio emission. As seen in the previous section, some
YSOs have been mistaken for PNe. We here examine the ASKAP radio
detections of a small sample of particularly well characterised YSOs from
the groundbreaking {\it Herschel Space Observatory} spectroscopy sample of
\cite{Oliveira2019}.

Of the seventeen YSOs (within the LMC), three had initially been associated with one
compact region of star formation and would be blended in our radio survey.
These were not detected; among the remainder, nine were detected (within a
search radius of 10 arcsec; all were found within $\leq$ 6 arcsec) including all three YSOs from the unique star-forming
region LHA\,120-N\,113 in which complex organic molecules were discovered
with the Atacama Large (sub-)Millimetre Array (ALMA) by \cite{Sewi_o_2018}. Table \ref{YSO_table} lists their properties. They span a range in radio
integrated flux densities from 0.45 to 47 mJy, and all but two are brighter than the PNe distribution
(Figure \ref{fig:PNe_cum_rad_lum_func}). Because they are among the more extremely luminous, massive
and consequently radio brighter YSOs they appear relatively isolated and
unresolved against the ambient emission, except for two clearly extended
sources both in the LHA\,120-N\,44 region.

\begin{table*}
\caption{ASKAP 888 MHz radio detections within 10 arcsec of sources from the {\it Herschel} spectroscopy of the YSO sample \citep{Oliveira2019}. The names are as in \citep{Oliveira2019}, the positions are those of the radio source, and the distance ($\Delta$) is that between the radio source and the YSO coordinates (IR-based).}
\begin{tabular}{lccccccccll}
\hline\hline
Name & EMU ID & RA (J2000) & Dec (J2000) & \multicolumn{2}{c}{----- peak -----} & \multicolumn{2}{c}{--- integrated ---} & $\Delta$ & Comments \\
 & & $^h$ $^m$ $^s$ & $^\circ$ $^\prime$ $^{\prime\prime}$ & \llap{$F$} (mJy) & \llap{$\sigma$} (mJy\rlap{)} & \llap{$F$} (mJy) & \llap{$\sigma$} (mJy\rlap{)} & $^{\prime\prime}$ & \\
\hline
IRAS\,04514$-$6931        & EMU ES J045111.7$-$692647 & 04 51 11.7 & $-$69 26 48 &  0.5\rlap{0} & 0.0\rlap{7} &  0.4\rlap{5} & 0.0\rlap{6} & 1.9 & \\
LHA\,120-N\,113\,YSO-1    & EMU EC J051317.6$-$692223 & 05 13 17.6 & $-$69 22 24 & \llap{3}9.0  & 0.6  & \llap{4}4.5  & 0.6  & 1.2 & \\
LHA\,120-N\,113\,YSO-4    & EMU EC J051321.4$-$692239 & 05 13 21.4 & $-$69 22 40 & \llap{4}0.7  & 0.6  & \llap{4}7.0  & 0.7  & 1.9 & \\
LHA\,120-N\,113\,YSO-3    & EMU EC J051324.7$-$692245 & 05 13 24.8 & $-$69 22 46 & \llap{1}1.2  & 0.6  & \llap{1}4.2  & 0.7  & 2.0 & \\
SAGE\,051351.5$-$672721.\rlap{9} & EMU ES J051351.4$-$672719 & 05 13 51.4 & $-$67 27 19 &  3.3\rlap{7} & 0.1\rlap{7} &  4.5\rlap{3} & 0.2\rlap{4} & 2.8 & \\
SAGE\,052202.7$-$674702.\rlap{1} & EMU ES J052202.7$-$674658 & 05 22 02.7 & $-$67 46 59 &  1.2\rlap{0} & 0.0\rlap{6} &  3.8  & 0.2  & 3.4 & extended \\
SAGE\,052212.6$-$675832.\rlap{4} & EMU EC J052213.3$-$675834 & 05 22 13.4 & $-$67 58 34 & \llap{3}0.2  & 1.1  & \llap{7}2.4  & 2.8  & 04.7 & extended \\
SAGE\,053054.2$-$683428.\rlap{3} & EMU ES J053054.5$-$683422 & 05 30 54.6 & $-$68 34 23 &  9.1\rlap{7} & 0.0\rlap{8} & \llap{1}4.9\rlap{0} & 0.1\rlap{3} & 6.1 & \\
ST\,01                    & EMU EC J053931.0$-$701216 & 05 39 31.0 & $-$70 12 16 &  2.4\rlap{3} & 0.0\rlap{8} &  3.1\rlap{8} & 0.1\rlap{1} & 1.1 & \\
\hline
\end{tabular}
\label{YSO_table}
\end{table*}

The most striking result is the sharp-cut division between those YSOs
detected with ASKAP and those not, in terms of the measurement of the gas
temperature based on far-IR emission lines of CO detected with {\it
Herschel} \citep[][their Table 6]{Oliveira2019}. All sources with
temperatures $T > 700$ K are detected, and all sources with $T < 700$ K are
not. This suggests a strong relation between the line emission from the
photo-dissociation regions at the interfaces of neutral and ionized gas,
and the free--free emission from the ionized regions surrounding the
nascent, presumed massive O-type stars. The brightest radio source,
LHA\,120-N\,113\,YSO-4 did not have a temperature determined from CO but
it is one of the two brightest [O\,{\sc iii}] 88 $\mu$m emitters \citep{Oliveira2019} and thus hosts a highly ionized compact or ultra-compact
H\,{\sc ii} region.


\subsection{Nov{\ae} and supernov{\ae}}

Of 46 nov{\ae} listed in Simbad within the ASKAP footprint, two are recovered
in our ASKAP catalogue (both in the G\textsc{old} list) within
$\sim2^{\prime\prime}$: Nova\,LMC\,1988\,b ($F_{888}=2.81\pm0.06$ mJy) and Nova
candidate LHA\,120-S\,162 ($F_{888}=0.46\pm0.05$ mJy). While the latter was
designated Nova\,LMC\,2001 it is a Galactic mid-M type
main-sequence emission-line star \citep{Morgan1992,Shafter2013}.
The former, Nova\,LMC\,1988\,b was a fast nova and the first neon nova
discovered in the LMC \citep{Sekiguchi1989}. This implies the eruption
occurred on a massive O--Ne--Mg white dwarf. Neon nov{\ae} are energetic and not rare, so it is not immediately clear why this should have been the only nova
detected with flux~density~$>$ mJy.

SN\,1987A is detected as a bright point source, with $F_{888}=1.1432\pm0.0013$ Jy. It is listed as EMU\,ES\,J053527.8$-$691611 in the B\textsc{ronze} source list. Its proximity to 30\,Doradus
prevented its original inclusion in the catalogue due to the complexity of
the surroundings even though SN\,1987A itself stands out clearly; it was
reinstated subsequently. The radio emission primarily arises from the
$1\rlap{.}^{\prime\prime}6$-diameter torus, which was resolved at 92 GHz
for the first time by \cite{Laki2011}. The integrated flux density at 888~MHz have continued to increase by $\sim27$ per cent since 2013--2014 when it was
$\approx0.90$ Jy \citep{Callingham2016}, but at the slower pace
observed after an initial exponential growth until 2009 \citep[day $\sim8000$, ][]{Ng2013}.

While radio detections are made at the locations of at least two dozen SNe in
background galaxies, in none of the cases could we unequivocally ascribe the
radio emission to the SN and in most cases the host galaxy dominates.


\subsection{Supergiants}

Various classes of massive supergiant stars could be detectable at low
radio frequencies, including Luminous Blue Variables (LBVs), Wolf--Rayet
(WR) stars -- especially if in colliding-wind binaries -- and Be
or B[e] stars (massive B-type stars showing permitted emission lines from
an excretion disk and/or forbidden emission lines from a more tenuous
envelope, respectively). 

Nine such sources from the far-IR study of \cite{vanLoon2010} were
pursued, resulting in the detection of four of these: (a) the dusty B[e] star
IRAS\,04530$-$6916 \citep{vanLoon2005} seen on the image as a point source with a peak flux density $\sim0.9$ mJy
\footnote{radio emission had been detected by \cite{Filipovic1995}} though it had not made the cut for the catalogue, (b) the
Of/Of?p--WR transition object Brey\,3 \citep[IRAS\,04537$-$6922;][]{Heydari1992} at $F_{888}=5.25\pm0.06$ mJy,
(c) emission-line object IRAS\,05047$-$6644 at $F_{888}=12.27\pm0.06$ mJy and
(d) IRAS\,05216$-$6753 at $F_{888}=3.77\pm0.11$ mJy -- likely a massive
early-type star illuminating a dusty envelope \citep{Chen2009}. It is
likely the nature of the 888 MHz radiation mechanism in these sources is
free--free emission from ionized circumstellar gas.

LBV S\,Dor may have been detected as a point source with a peak flux density $\sim0.19$ mJy, but it is blended with
much brighter adjacent emission from the N\,119 H\,{\sc ii} region.
The radio emission near R\,71, at $\sim$ 12 arcsec distance ($\sim
3$ pc at the distance of the LMC) to the South--West in an otherwise
sparse field, is unresolved and peaks at $\sim0.21$ mJy. R\,71 had also not been detected within a $3\sigma$ level of 24 $\mu$Jy at 5 GHz in 2015 \citep{Mehner2017}.
While no circumstellar nebula appears to be present around R\,71 (based on HST H$\alpha$ imaging before the recent eruption of R\,71), there is a hint for a very small (less than 1 arcsec or 0.25 pc) circumstellar nebula around S\,Dor due to the [N\,{\sc ii}] emission lines detected in a long-slit \'echelle spectrum \citep[][]{2003Weis}. Unfortunately, the possible nebula is not detected in HST H$\alpha$ imaging \citep[][]{2003Weis}.

LBV nebul\ae\ have been detected at higher radio frequencies before, for instance, R\,127, R\,143, S\,61 and S\,119 \citep{Agliozzo2012,Agliozzo2017a,Agliozzo2017b,Agliozzo2019}. Cross-matching these sources with the ASKAP LMC 888 MHz catalogue revealed integrated fluxes for R\,127 (EMU\,ES\,J053643.4$-$692947) of $F_{888}=2.69\pm0.18$ mJy, S\,61 (EMU\,EC\,J054551.8$-$671426) of $F_{888}=2.31\pm0.06$ mJy and S\,119 (EMU\,ES\,J053125.5$-$690537) of $F_{888}=1.09\pm0.11$ mJy, all point-like sources. 

\subsection{X-ray binaries}

Simbad lists 66 X-ray binaries within the ASKAP footprint. With the only exception of LMC\,X-2 (a low-mass X-ray binary), these  are high-mass X-ray binary (HMXB) systems formed by a compact object (in most cases a neutron star) accreting matter from a massive companion star \citep{2016MNRAS.459..528A}. 
From about 20 systems X-ray pulsations were discovered, which identify the spin period of the neutron star.
In many of the others cases the nature of the X-ray source is less clear; low statistical data quality and/or uncertain X-ray position do not allow identification of the optical counterpart.

Four of the 66 (candidate) X-ray binaries were successfully recovered in the 888 MHz catalogue -- three in the G\textsc{old}
and one in the S\textsc{ilver} source list. They are all well detected, coincident and isolated in the radio.

The radio-brightest of the four at $F_{888}=9.15\pm0.05$ mJy correlates with RX\,J0457.2$-$6612, a little-known X-ray source, listed in the ROSAT PSPC X-ray catalogue of the LMC region by \cite{Haberl1999A} and proposed by \citet{2002A&A...388..100K} as X-ray binary.
However, there is no star within 20$^{\prime\prime}$ brighter than $V=18.8$ mag, which most likely excludes an HMXB nature.
The X-ray source is probably identified with a QSO listed in \citet{2019MNRAS.490.5615B}, which is further supported by the ASKAP radio detection.

The second brightest (EMU\,ES\,J050123.5$-$703329, $F_{888}=3.25\pm0.05$ mJy) coincides with \textit{Einstein} source Cal\,9 \citep{1984ApJ...286..196C}, which was identified with ROSAT source RX\,J0501.6$-$7034 by \citet{1999AJ....117..927S}.
The authors list as possible optical counterpart HV\,2289 -- an
early-B type supergiant with an orbital period of $P=6.94$ d \citep{Bianchi1984}. X-ray positions from Einstein and ROSAT are not precise enough to allow a secure optical identification, 
but the source was also detected by XMM--{\it Newton} \citep[4XMM\,J050123.3$-$703333 in the 4XMM-DR9 catalogue; ][]{2020A&A...641A.136W} with a position error of $1\rlap{.}^{\prime\prime}5$ (1$\sigma$). 
This position is $2\rlap{.}^{\prime\prime}4$ from the Gaia DR2 position \citep{2018A&A...616A...1G} of HV\,2289, consistent within the uncertainty. Also the radio position is consistent within errors with the 4XMM source.
\citet{2002A&A...385..517N} give B0\,Ve for the spectral type, which suggests a classical Be/X-ray binary instead of a supergiant system. On the other hand, the orbital period of 6.94 d would be too short for a Be/X-ray binary 
\citep[e.g., ][]{2016A&A...586A..81H} and favour a supergiant system.
The radio detection of this system makes it even more puzzling.

The two faintest radio detections are associated with  RX\,J0535.8$-$6530 ($F_{888}=0.57\pm0.05$ mJy) and
[SG2005]\,LMC\,15 ($0.40\pm0.05$ mJy). 
The former was found to be a variable X-ray source by \cite{Haberl1999B}; an optical counterpart was suggested to be a star by \cite{Grazian2002}, and it is also detected
with {\it Spitzer} and WISE.
The GLADE catalogue \citep{2018MNRAS.479.2374D} and \citet{2019MNRAS.490.5615B} list a background
galaxy with redshift 0.05. The radio detection supports this identification.

The latter, [SG2005]\,LMC\,15 was proposed as an HMXB candidate by \cite{Shtykovskiy2005}, but the optical counterpart is not coincident with the XMM--{\it Newton} and
{\it Spitzer} source and the real counterpart of the X-ray and radio source is likely the QSO MQS\,J054433.81$-$682813.7 given in the Million Quasars catalogue \citep[Milliquas V6.3, ][]{2015PASA...32...10F}.

\begin{figure}

	\includegraphics[width=\columnwidth]{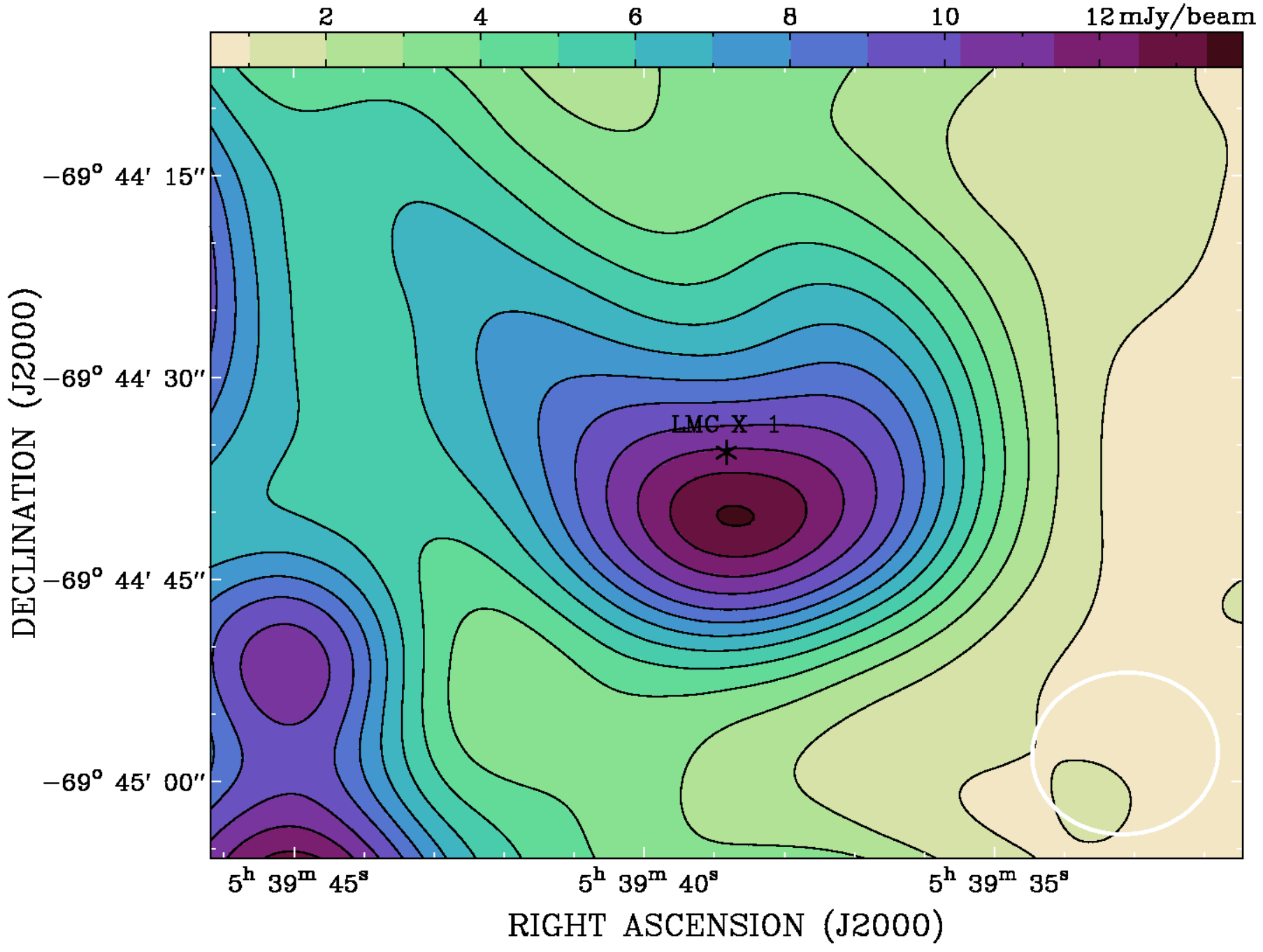}

    \caption{X-ray binary LMC\,X-1 seen in the focus of a radio arc in the 888 MHz ASKAP LMC image. The beam size for ASKAP 888 MHz is $\sim13\rlap{.}^{\prime\prime}9\times12\rlap{.}^{\prime\prime}1$ and is represented by the white ellipse in the bottom right.}
    \label{fig:LMC_X1}
\end{figure}

The most famous X-ray binaries in the LMC, LMC\,X-1 (a black hole + supergiant system) and LMC\,X-4 (a neutron star accreting from a Roche-lobe filling supergiant) 
were not detected.
However, we do detect a radio arc with LMC\,X-1 in its focus
(Figure \ref{fig:LMC_X1}), peaking at $4^{\prime\prime}$ due South (1 pc in projection at the distance of the LMC). 
If this is a bow shock \citep{Hyde2017} ahead of the supersonic motion of LMC\,X-1 then this means the binary must have originated from {\em outside} the prominent star-forming region LHA\,120-N\,159 into which it would seem to be heading. The system may have run away from N\,159 and subsequently received a lateral kick when the supernova created the black hole.

\subsection{Galactic active stars}

Galactic stars seen in the direction of the Magellanic Clouds are at high
Galactic latitude and radio-loud stars among them are most commonly nearby
($\sim$100 pc) flare stars (M-dwarves) or interacting tight binaries.

\begin{figure}

	\includegraphics[width=\columnwidth]{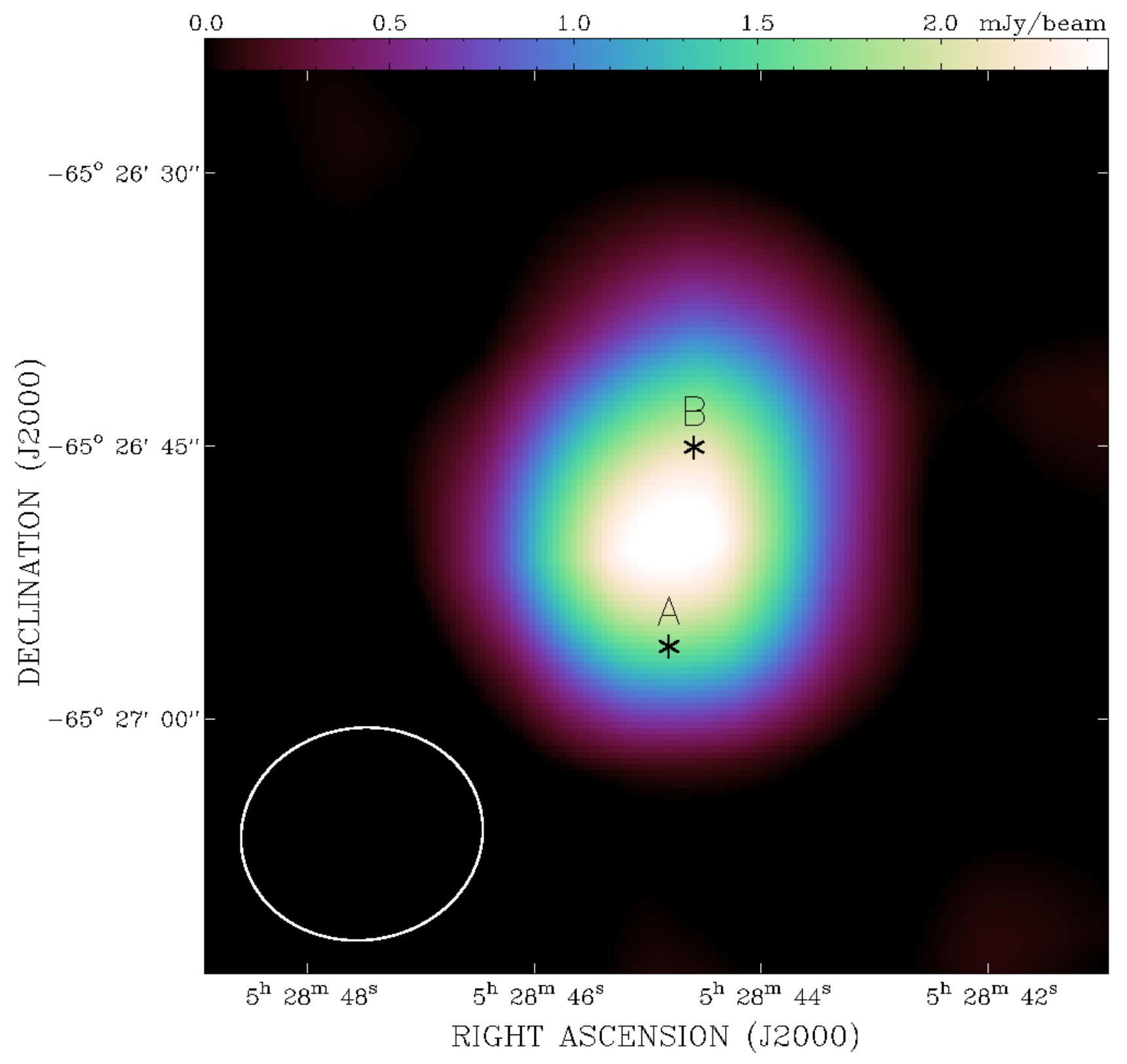}

    \caption{LMC AB\,Dor, a hierarchical young active system detected in the 888 MHz ASKAP LMC image where emission coming from the main component and the B component contribute. ASKAP 888 MHz beam size is $\sim13\rlap{.}^{\prime\prime}9\times12\rlap{.}^{\prime\prime}1$, represented by the white ellipse in the bottom left.}
    \label{fig:LMC_ABDor}
\end{figure}

We detect the hierarchical young active system AB\,Dor \citep{Pakull1981, Collier1988,Wolter2014,Azulay2015,Azulay2017} and resolve the emission coming from the main (K-type) component and
the B component (Figure \ref{fig:LMC_ABDor}) -- the latter of which itself is a
spectroscopic M-dwarf binary. \cite{White1988} had ruled out 8-GHz
emission coming from component B, but the system was later resolved at 5
GHz by \cite{Beasley1993} and \cite{Lim1993}, with the latter work showing the
highly polarized and variable nature of component B. The system has
another component, C, which is a late-M dwarf with a sub-stellar companion
\citep{Climent_2019} close to the main component but on the side of
component B, so we cannot discard the possibility that some of the radio
emission originates from it. The ASKAP emission is heavily blended, with a
peak and integrated flux density of $F_{888,{\rm peak}}=2.80\pm0.05$ mJy
and $F_{888,{\rm total}}=3.57\pm0.07$ mJy, respectively. \cite{Slee1984} detected it at 5 GHz (5.2 mJy); \cite{Vaughan1986} subsequently detected it at 843 MHz at a similar level (4 mJy) and
our measurement confirms the flat spectrum (in $F_\nu$). \cite{Collier1982}
characterised its activity as that of an FK\,Com system, which is
associated with rapid rotation and evolution off of the main sequence \citep{Oliveira1999}.
\cite{Rucinksi1983}, however, argued for youth and a post-T\,Tauri status and
their detection of lithium seemed to prove that. \cite{Innis1986} \citep[see also][]{Ortega2007} determined the association of the
AB\,Dor system with the Pleiades moving group \citep[later revisited as the AB\,Dor moving group -- see, e.g., ][]{Barenfeld2013}, confirming its
relative youth. The activity of AB\,Dor is usually associated with
coronal, magnetic activity \citep[e.g., ][]{Donati1997,Donati1999,Jardine2002,Sanzforcada2003}. AB\,Dor is at the same distance of 15 pc as the eruptive pair of M dwarves CD\,$-$38$^\circ$11343, which was detected with ASKAP by \cite{Riggi2021} at essentially the same integrated radio flux densities (at 912~MHz) as AB\,Dor.

The RS\,CVn system AE\,Men is detected, at $F_{888}=1.51\pm0.06$~mJy.
At a Gaia distance of $d=288$ pc this corresponds to a radio luminosity of
$2\times10^{17}$ erg s$^{-1}$ Hz$^{-1}$, placing it among the top 10 per cent integrated radio flux densities of this kind of system \citep{Morris1988, Gudel2002}. This system is of K2\,III+F/G spectral type and it
is suggested to exhibit non-thermal (gyrosynchrotron) emission. What makes this object unusual compared to the
$\sim100$+ known population of RS\,CVn type systems is its X-ray (ROSAT),
H$\alpha$ and especially IR (AllWISE) detection. AE\,Men was observed on
several occasions with MOST and ATCA \citep{Slee1987,Vaughan1987,Wendker1995} but this is its first radio continuum
detection. We reprocessed the ATCA observations from 1994 and confirm that it was not
detected.

The $\beta$\,Lyr{\ae}-type eclipsing binary BK\,Dor (= CD\,$-67^\circ435$)
is also detected for the first time at radio frequencies, with
$F_{888}=0.90\pm0.07$ mJy. With the Gaia distance of 279 pc this yields a
radio luminosity of $8\times10^{16}$ erg s$^{-1}$ Hz$^{-1}$, not as high
as AE\,Men but still luminous. It was discovered as a variable X-ray source by
\cite{Fuhrmeister2003} before being associated with the G7\,III
star by \cite{Torres2006} and found to be a short-period ($P=2.2$ d)
eclipsing binary by \cite{Szczygiel2008}. In such a system the two
stars are highly distorted and exchange mass.

\section{Background sources} \label{Backgroundsources}

Here we discuss the extragalactic sources seen behind the LMC, which dominate the source counts. The radio emission from these sources is due to synchrotron radiation from relativistic electrons in magnetic fields and free--free emission from H\,{\sc ii} regions.

\subsection{Radio galaxies}

\begin{figure*} 
\centering
\begin{tabular}{c}
	\includegraphics[trim={0 0.7cm 0 0.6cm},width=\textwidth]{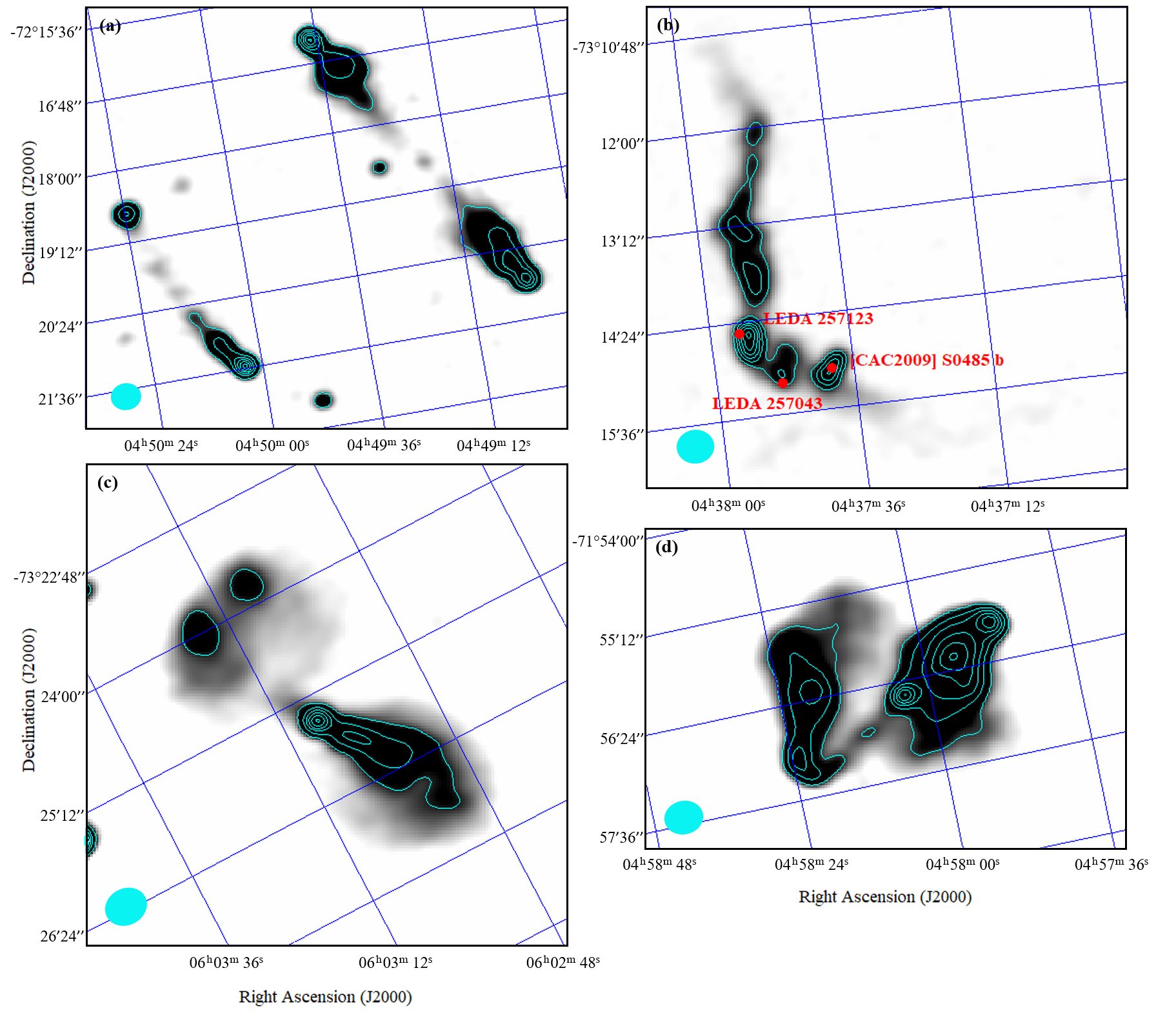}
	\end{tabular}
    \caption{Four examples of radio emission from galaxies. The structure of the four radio sources is shown by the cyan contours that are drawn at equally spaced values in brightness. The beam size for ASKAP 888 MHz is $\sim13\rlap{.}^{\prime\prime}9\times12\rlap{.}^{\prime\prime}1$, which is represented by the cyan ellipse. (a) Two examples of jets originating from a less bright central source (FR II systems) where contours span from 0.2--29 mJy beam$^{-1}$. (b) Galaxy cluster at $z\sim$ 0.06. Low surface brightness at lower right is 0.25--0.45 mJy beam$^{-1}$. The contours span from 0.2--19 mJy beam$^{-1}$. (c) A jetted system that reveals interesting morphology in the jets. Contours span from 3.7--28 mJy beam$^{-1}$. (d) A central radio source from which two jets originate. These jets are bent as they come into contact with the intra-cluster medium. The contours span from 3--19 mJy beam$^{-1}$.}
    \label{fig:galaxies}
\end{figure*}

The ASKAP LMC image features some spectacular radio galaxies, showcasing a clearer morphology of these sources than has been seen before. Figure \ref{fig:galaxies} shows a few examples of these radio systems.

\textbf{Figure \ref{fig:galaxies}, (a)}. The NW quadrant shows a Fanaroff--Riley II (FR II) system \citep{FaRi1974}, which are luminous radio systems brighter near their extremities than near their centres. This is a known radio source, PMN\,J0449$-$7219 \citep{Filipovic1998} and is seen as three islands (islands 14012, 13945 and 13724) in our ASKAP catalogue. The jets originate from galaxy WISEA\,J044917.77$-$721843.7 (a.k.a.\ 2MASS-6X\,J04491744$-$7218434, VISTA\,J044917.454$-$721843.85) for which \cite{Bilicki2016} give an extinction-corrected R-band magnitude of 19.62 and a photometric redshift of 0.47. With a radio size of $5\rlap{.}^{\prime}5$, this makes it a giant radio galaxy of projected linear size of $\sim$1.9 Mpc. A similar source in the SE corner of the image is most likely
hosted by the galaxy VISTA\,J045013.036$-$722036.25 ($K_{\rm sp}=15.45$ mag, a.k.a.\
WISEA\,J045013.00$-$722036.0, 2MASS-6X\,J04501307$-$7220362), for
which \cite{Bilicki2016} give an extinction-corrected 
R-band magnitude of 19.68 and a photometric redshift of 0.35. With a radio size of $3\rlap{.}^{\prime}4$, this makes it another giant radio
galaxy of projected linear size of $\sim$1.0 Mpc. Whilst the radio emission  from the ends of the jets can be seen in the SUMSS 843~MHz image, the fainter, central part of this source, where the radio jets most likely originate, has not been seen before.

\textbf{Figure \ref{fig:galaxies}, (b)}. This is likely a wide-angle tailed (WAT) radio galaxy hosted
by LEDA\,257123 (EMU\,EC\,J043753.4$-$731437), one of the members of galaxy cluster Abell\,S0485 
with a mean spectroscopic redshift of 0.061 based on seven galaxies 
from the 6dF redshift survey \citep{Jones2009}. Neither
the host galaxy of the WAT nor LEDA\,257043 (EMU\,EC\,J043740.1$-$731508) or [CAC2009]\,S0485\,b (a.k.a.\ 2MASX\,J04373976$-$7315056), apparently
superposed on the SW tail of this WAT, have a spectroscopic redshift,
and the host galaxy LEDA\,257123 is not among the three brightest galaxies
\citep{Coziol2009}. Assuming the cluster redshift for the host,
the radio size of the WAT of $6\rlap{.}^{\prime}2$ corresponds to a projected linear
size of 440 kpc.

\textbf{Figure \ref{fig:galaxies}, (c)}. ASKAP 888-MHz image reveals that this known radio source
(PMN\,J0603$-$7325) has an S-shaped morphology indicating precessing jets. The strong radio core is coincident with the galaxy 2MASX\,J06030694$-$732529 (EMU\,EC\,J060306.4$-$732529)
for which \cite{Bilicki2016} give a corrected R-band magnitude of 
16.32 and a photometric redshift of 0.156. With an angular extent of $3\rlap{.}^{\prime}5$ this corresponds to a projected linear size of 570 kpc.

\textbf{Figure \ref{fig:galaxies}, (d)}. This radio source (PKS\,0458$-$720, PMN\,J0458$-$7156, EMU\,EC\,J045804.2$-$715635) shows an even more
extreme S-shape likely due to projection along the line of sight. 
Its radio core coincides with 2MASX\,J04580433$-$7156347 \citep[a.k.a.\ SMSS\,J045804.30$-$715635.0, $r_{\rm Petro}=17.07$ mag, ][]{Wolf2018}
, but no (photometric) redshift is available. 

These galaxies provide examples of the potential for exploring the morphologies in greater detail than previously, and for locating the host galaxies of these extragalactic marvels. We provide more in-depth analysis using high-resolution IR images and optical spectroscopy in a followup paper (Pennock et al., in prep.).

\subsection{Comparison with the GLEAM 4-Jy sample}
\begin{figure*} 
\centering
\begin{tabular}{c}
	\includegraphics[width=0.87\textwidth]{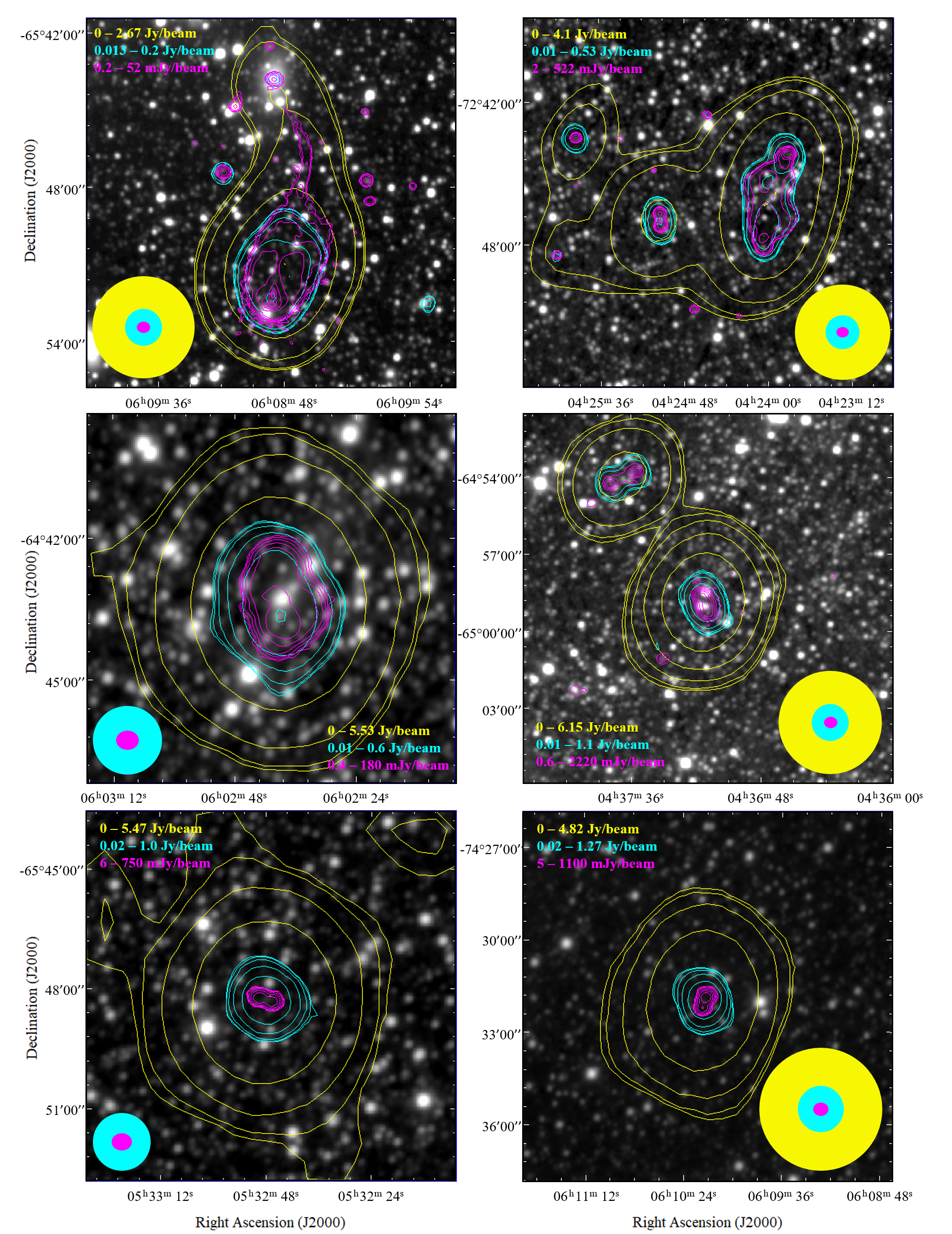}
	\end{tabular}
    \caption{Comparison between ASKAP (magenta contours), SUMSS (cyan contours) and GLEAM (yellow contours) that represent logarithmically spaced brightness levels in the ASKAP image, the minimum and maximum of which are annotated in their respective colour. The background image is the WISE W3 band. The beam size for ASKAP 888 MHz is $\sim13\rlap{.}^{\prime\prime}9\times12\rlap{.}^{\prime\prime}1$ and is represented by the magenta ellipse. The beam size for SUMSS 843 MHz is $\sim43^{\prime\prime}\times43^{\prime\prime}$ and is represented by the cyan circle. The beam size for GLEAM 72--231 MHz is $\sim2^\prime\times2^\prime$ and is represented by the yellow circle. The sources are (top-left) G4Jy 595, GLEAM\,J060849$-$655110, (top-right) G4Jy 453, GLEAM\,J042358$-$724601, (middle-left) G4Jy 587, GLEAM\,J060239$-$644324, (middle-right) G4Jy 482, GLEAM\,J043709$-$645853, (lower-left) G4Jy 553, GLEAM\,J053248$-$634820 and (lower-right) G4Jy 596, GLEAM\,J061014$-$743159.}
    \label{fig:GLEAM1}
\end{figure*}

The GaLactic and Extragalactic All-sky MWA 4-Jy  sample \citep[G4Jy; ][]{White2020b,White2020a}, from the Murchison Widefield Array (MWA), provides 
flux-densities across the range 72 to 231 MHz at DEC $<30^\circ$). G4Jy  is a complete sample of the brightest radio sources, $F > 4$ Jy, which are expected to be mainly AGN with powerful radio-jets (strong optically-thin synchrotron emission). Six of the sources from this sample lie within the ASKAP LMC field, all of which were detected in the ASKAP observations. These are shown in Figure \ref{fig:GLEAM1} and show the large improvement the ASKAP image offers in resolution, which allows the identification of the optical/infrared counterpart of these radio sources, and while almost all host galaxies are identified in the WISE image, some are very faint. These GLEAM sources, which appear at first glance to be one source, are revealed to be multiple sources by ASKAP. The majority of these sources have the appearance of FR II sources. For the sources G4Jy 482, G4Jy 553 and G4Jy 596, ASKAP shows that they are double sources not resolved in SUMSS.

\textbf{Figure \ref{fig:GLEAM1}, top-left.} Contours of G4Jy 595 (a.k.a.\ GLEAM\,J060849$-$655110) from ASKAP reveal that the tear drop shaped radio source seen in GLEAM is in fact two sources. The 
southern source is an extreme bent tailed source hosted by the galaxy
2MASX\,J06085498$-$6552559 (a.k.a.\ WISEA\,J060854.94$-$655255.6) with $z=0.03752$ \citep{Jones2009}. The northern one is a radio point source hosted by the galaxy ESO\,086-G062 
(a.k.a.\ WISEA\,J060852.00$-$654350.2), with $z=0.037$ 
\citep{Wegner2003}, which appears to be part of a galaxy group that the bent tailed radio source has moved through. In fact, \cite{Saulder2016} lists the group of three galaxies 2MRS\,3198 at $z=0.03774$. The third member galaxy is 2MASX\,J06090657$-$6544539 (a.k.a.\ LEDA\,310538) with $z=0.03828$ \citep{Huchra2012}, which lies $\sim2^\prime$ SE of ESO\,086-G062 and is also detected as a radio point source.

\textbf{Figure \ref{fig:GLEAM1}, top-right.} ASKAP reveals that G4Jy 453 (GLEAM\,J042358$-$724601) comprises multiple sources. The western-most ASKAP radio source features an X-shaped radio galaxy known as PKS\,0424$-$728 at a redshift of $\sim$0.19 \citep[inferred from photometry;][]{Subrahmanyan1996}. With more recent photometric redshifts of 0.229 and 0.214 from \cite{Bilicki2014} and \cite{Bilicki2016} its angular size of $4\rlap{.}^\prime3$ gives it a projected linear size of 880 kpc. An unresolved source at 888 MHz is hosted by WISEA\,J042434.02$-$724241.3 with an even fainter 888 MHz source $\sim24^{\prime\prime}$ SW with no optical/IR counterpart and the X-ray source 1RXS\,J042432.7$-$724253 in between them. The central source is revealed as a double radio source of size $\sim33^{\prime\prime}$ hosted by WISEA\,J042501.60$-$724700.3 with no optical counterpart. Two ASKAP point sources at the E edge of the radio complex can be identified with WISEA\,J042549.29$-$724330.8 (NE) and WISEA\,J042600.47$-$724828.9 (SE).

\textbf{Figure \ref{fig:GLEAM1}, middle-left.} This is G4Jy 587 a.k.a.\ PKS\,0602$-$647 or GLEAM\,J060239$-$644324. At the centre of this radio structure is a pair of galaxies, the brighter of which, LEDA\,319866 with $z=0.045$ \citep{Burgess2006}, is likely the host. Nearby galaxies with known redshifts show similar redshifts, such as the galaxy NW of the host that lies on the edge of the ASKAP contours, which is 2MASX\,J06023279$-$6442344 with $z\sim0.048$ \citep[photometric;][]{Bilicki2014}.

\textbf{Figure \ref{fig:GLEAM1}, middle-right.} The central source is G4Jy 482 (GLEAM\,J043709$-$645853) and the NE source is GLEAM\,J043738$-$645400 (not in the G4Jy sample). The former is MRC\,0436$-$650 a.k.a.\ PKS\,0436$-$650 hosted by the galaxy DES\,J043708.39$-$645901.9 \citep[$r=17.97$ mag, $z_{\rm sp}=0.360$,][]{Burgess2006}. The NE double source is hosted by the QSO candidate DES\,J043738.82$-$645402.1 \citep[$r =21.53$ mag, no $z_{\rm phot}$ available,][]{Abbott2018} a.k.a.\ CWISE\,J043738.83$-$645402.0 \citep{Marocco2021}. The source $\sim2\rlap{.}^\prime7$ SE of G4Jy 482 is DES\,J043724.19$-$650105.6 a.k.a.\ WISEA\,J043724.19$-$650104.9 with high stellarity, i.e.\ likely a QSO.

\textbf{Figure \ref{fig:GLEAM1}, lower-left panel.} The central source is G4Jy 553 (GLEAM\,J053248$-$634820), also known as PKS\,0532$-$638 and MRC\,0532$-$638 or ATPMN\,J053246.7$-$634820 \citep{McConnell2012}, hosted by the QSO candidate SMSS\,J053247.95$-$634818.1 \citep[$g_{\rm Petro} = 19.89$ mag,][]{Wolf2018} a.k.a.\ WISEA\,J053247.87$-$634816.7
with typical QSO colors ($W12\sim1.0$ mag, W$23\sim3.0$ mag), suggesting a redshift of $z>1$ \citep[see e.g.,\ Fig.\ 2 of][]{Krogager2018}.

\textbf{Figure \ref{fig:GLEAM1}, lower-right panel.} This is G4Jy 596 (GLEAM\,J061014$-$743159, PKS\,0611$-$745), which was identified by \cite{White2020b} with WISEA\,J061013.90$-$743201.6 and is not detected in the optical.


\subsection{IR properties of radio sources}
The AllWISE catalogue \citep{Cutri2013} is a well-explored all-sky survey with photometry from 3.35 to 22.1 $\mu$m. This wavelength regime allows for a selection of both obscured and unobscured AGN, which makes it a great catalogue to cross-match to discover extragalactic sources. Selection criteria for different types of AGN/galaxies are well documented \citep[e.g., ][]{Wright2010,Jarrett2011,Mateos2012,Stern2012,Assef2013,Nikutta2014} and the criteria from \cite{Wright2010} will be used here to discuss the radio population.

Cross-matching the AllWISE catalogue with the new ASKAP catalogue with a search radius of 5 arcsec and a maximum magnitude error in the W1 (3.4 $\mu$m), W2 (4.6 $\mu$m) and W3 (12 $\mu$m) bands of 0.2 mag yielded a total of 14,333 sources (combined G\textsc{old} and S\textsc{ilver}, which contain 7,413 and 6,920 sources, respectively). This search radius was used as the density of coincidences drops off sharply, approaching
that of chance coincidences (tested by applying a large offset in RA and DEC and then cross-matching again) beyond a
search radius of 5 arcsec, and coincidences found between
6--10 arcsec of stellar sources (including PNe) almost
invariably turned out to be implausible associations.
As a comparison with ASKAP detected sources, the sources within the LMC field that had no ASKAP detection (no ASKAP emission above $5\sigma$ of the local RMS) within 5 arcsec were considered. This amounted to 253,533 AllWISE non-ASKAP detected sources.

\begin{figure*}

	\includegraphics[trim={0.5cm 1cm 1cm 0.5cm},width=\textwidth]{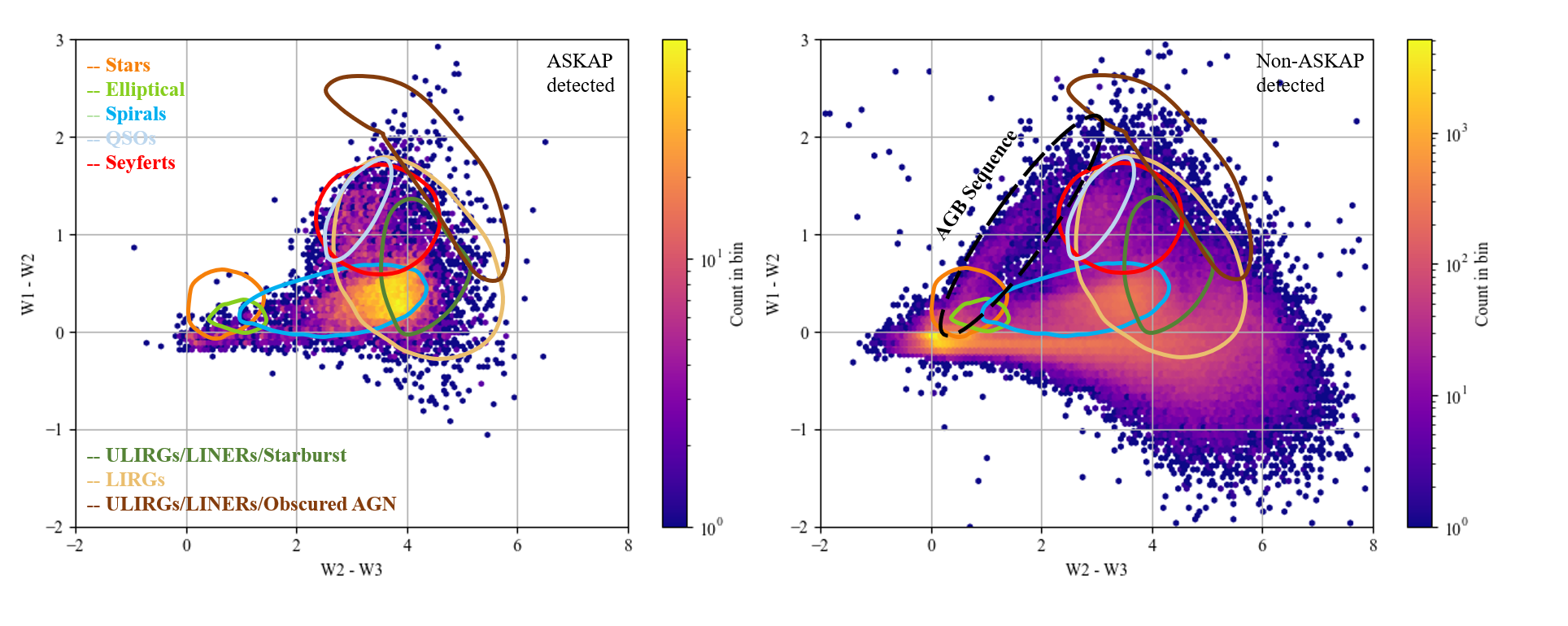}

    \caption{(Left) LMC AllWISE colour--colour diagram of the combined G\textsc{old} and S\textsc{ilver} source lists. (Right) LMC AllWISE colour--colour diagram of the non-ASKAP detected sources in the LMC field.}
    \label{fig:LMC_WISEcolour}
\end{figure*}

We plotted the ASKAP detections and non-detections on the W1--W2 vs.\ W2--W3 colour--colour diagram (Figure \ref{fig:LMC_WISEcolour}),  based on Figure 12 of \cite{Wright2010}. The majority of the radio detections are extragalactic sources. The stellar AGB sequence seen in the non-ASKAP detected diagram (right) is clearly not visible in the ASKAP detected diagram (left). The B\textsc{ronze} source list also shows a similar colour--colour diagram to the combined G\textsc{old} and S\textsc{ilver} source lists and therefore the majority of sources are extragalactic. 

Where LMC ISM dust emission affects the WISE detections it would also be
too bright at radio frequencies. Extinction caused by dust within the LMC
would affect the optical brightness of background AGN and galaxies. The
vast majority of the ASKAP field is affected only by low or modest levels
of such extinction ($A_V\ll 1$~mag), judging from extinction maps such as
those produced by \cite{Skowron2020}, which would not cause major
difficulties for optical spectroscopy and certainly not lead to any bias
between the extragalactic populations with and without such spectroscopy.

\subsubsection{Spectroscopic AGN}
There are 657 spectroscopically observed AGN, from the milliquas catalogue of \citet{Flesch2019cat,Flesch2019pap}, in the field of the ASKAP LMC image. The highest contributions are from \cite{Kozlowski2012,Kozlowski2013}, \cite{Geha2003}, \cite{Esquej2013} and \cite{Ivanov2016}, contributing 547, 24, 23 and 10 objects, respectively.

This list of spectroscopically observed AGN was cross-matched with the combined G\textsc{old} and S\textsc{ilver} ASKAP LMC source lists with a search radius of 5 arcsec. This gave a total of 190 known AGN with radio detections, only 14 ($<8$ per cent) of which had known radio detections prior to the present ASKAP observations. As expected, all known radio associated AGN are in the new ASKAP catalogue. The classifications of the known AGN that were radio detected and non-radio detected are described in table \ref{tab:knownAGNclass}.

\begin{table}
\caption{Classifications of known AGN that have been detected with ASKAP and those that have not. Q = QSO, type-I broad-line core-dominated; A = AGN, type-I Seyfert/host-dominated; B = BL Lac object; N = narrow-line AGN (NLAGN), type-II Seyfert/host-dominated (includes unknown number of legacy narrow emission line galaxies (NELGs)/emission line galaxies (ELGs)/
low-ionization nuclear emission-line region (LINERs)); R = radio associated ; X = X-ray associated.}
    \centering
    \begin{tabular}{|l|c|c|c|c|c|c|c|}
        \hline\hline 
         & Total & A & Q & N & B & R & X\\
       \hline 
       Radio  & 202 & 67 & 117 & 5 & 1 & 12 & \mbox{~~}48\\
       Non-radio & 453 & 80 & 368 & 5 & 0 & \mbox{~~}0 & 110\\
       \hline
    \end{tabular}
    
    \label{tab:knownAGNclass}
\end{table}

The spectroscopically observed AGN have a selection bias. All had been chosen first and foremost because they are optically bright enough to take a spectrum of, which leads to a surplus of lower redshift AGN. Whether they are bright enough in the optical and a candidate AGN or not is determined from photometric surveys. The AGN has to dominate over the host to be selected as an AGN candidate, based on colour selections. If selected in the optical, where optical emission from AGN comes from the accretion disk and the broad line region (BLR), which introduces a bias towards unobscured AGN, typically type 1 AGN. IR provides a selection that is sensitive to both obscured and unobscured AGN, providing a more isotropic selection than in the optical. In this regime the dust obscuring the central AGN re-emits the absorbed optical emission into the IR. Most of the AGN behind the LMC were spectroscopically observed in the optical, even if they were selected in the IR, which lends the bias towards the more unobscured AGN yet again.

\begin{figure*} 
\centering
\begin{tabular}{c}
	\includegraphics[trim={1cm 1cm 0 0.7cm},width=\textwidth]{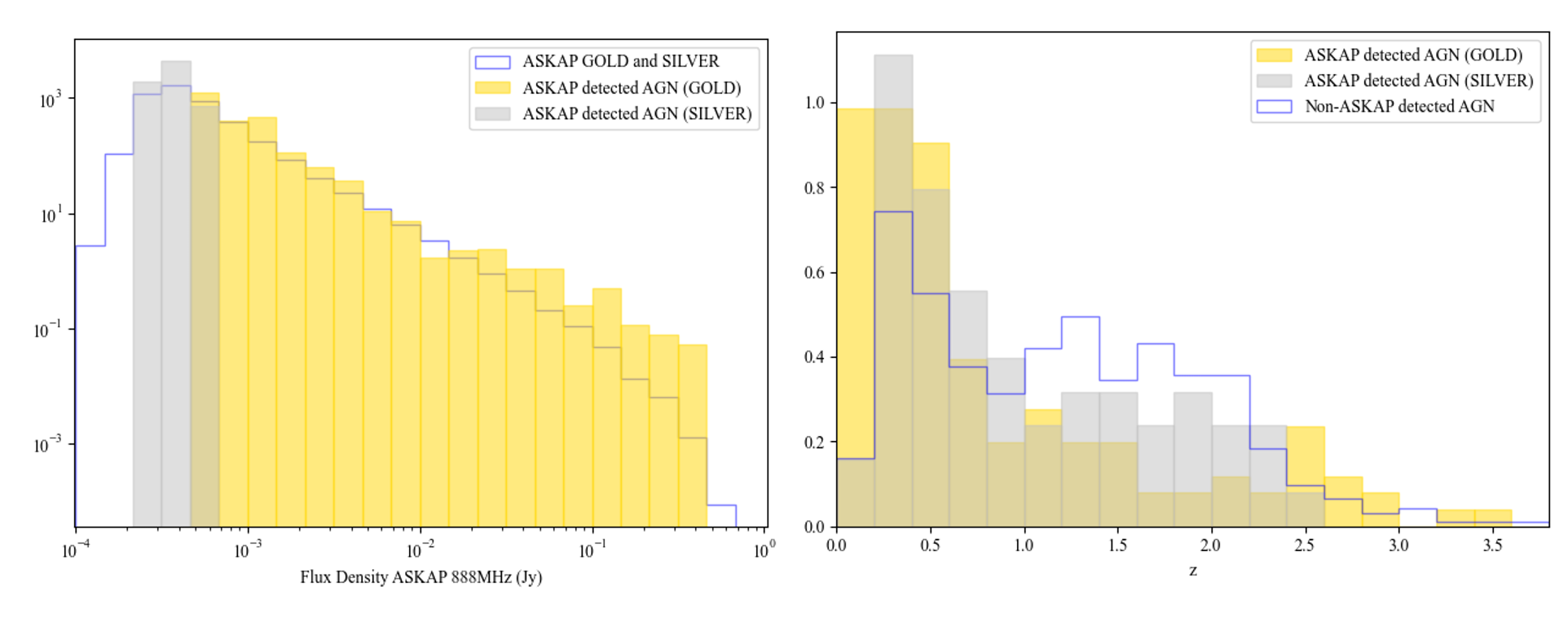}
	\end{tabular}
    \caption{(Left) Normalised histogram comparing integrated flux densities of spectroscopically observed radio AGN vs.\ all radio AGN in the LMC field. This shows that the spectroscopically observed radio AGN tend to brighter flux densities than the overall radio population. (Right) Normalised histogram comparing the redshift of spectroscopically observed radio AGN (separated into G\textsc{old} and S\textsc{ilver}) vs.\ all spectroscopically observed AGN in the LMC field. This shows that radio-detected spectroscopically observed AGN tend more towards lower redshift compared to non-radio detected spectroscopically observed AGN.}
    \label{fig:specradiocomp}
\end{figure*}

 The redshifts of the known radio AGN range from $z =$ 0.001--3.46. The distribution of integrated radio flux densities for spectroscopically observed radio AGN compared to all radio sources (shown in Figure \ref{fig:specradiocomp} (left)) reveals only a slight bias towards the brightest radio sources having more spectroscopic follow-up, which is not unexpected, since the radio brightness would have made them targets for spectroscopy. The slope of the histogram of flux density of known AGN detected is similar to the slope of the combined G\textsc{old} and S\textsc{ilver} source lists; below $\sim 10$ mJy the shape of the flux density functions are indistinguishable.
 
 The redshift distributions for radio-bright, radio-faint and non-radio sources (Figure \ref{fig:specradiocomp} (right)) are similar, with a slight bias towards lower redshift AGN being radio sources. While one might expect a bias towards higher redshift for radio sources compared to the overall distribution of galaxies, observational bias in the spectroscopy most likely negates that. The S\textsc{ilver} source list ($F_{888} < 0.5$~mJy) overall redshift distributions seem to be half way between the G\textsc{old} source list ($F_{888} > 0.5$ mJy) distribution and the non-radio distribution. Non-radio detected spectroscopically observed AGN are
generally at higher redshift, this could be because at higher redshift the optical
will in fact be the UV restframe, thus making AGN brighter and easier to
obtain an optical spectrum of. Furthermore, at higher redshift one gets a higher fraction of radio-quiet AGN, which will be detectable in the optical but too faint for ASKAP.

\begin{figure} 
\centering
	\includegraphics[trim={0.25cm 1cm 0.25cm 0.5cm},width=\columnwidth]{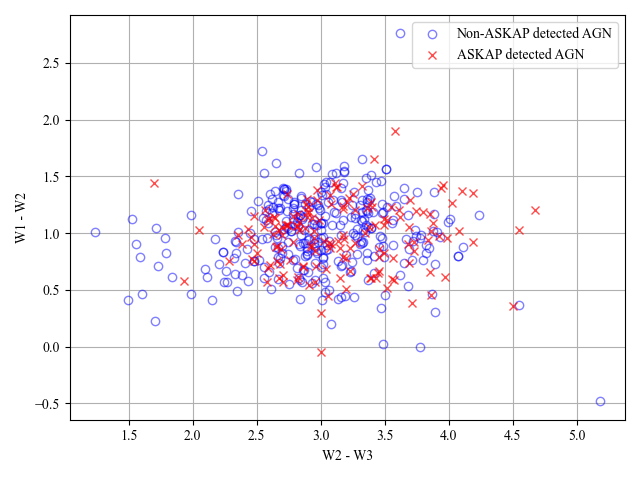}
    \caption{AllWISE colour--colour diagram of spectroscopically observed AGN, where those detected in the radio with ASKAP (red crosses) are compared with those not detected with ASKAP (blue circles). }
    \label{fig:AWspecradiocomp}
\end{figure}

Figure \ref{fig:AWspecradiocomp} compares, using AllWISE colours, the radio and non-radio detected AGN. Radio detected AGN overall appear redder in W2--W3 than those not detected in radio. A Kolmogorov--Smirnov test of the radio and non-radio detected populations in the W2--W3 colour band shows that the radio detected population is redder at a significance level of 10 per cent \footnote{the probability of the null hypothesis in the Kolmogorov--Smirnov test that these two samples are drawn from the same parent sample.}.

\begin{figure*} 
\centering
\begin{tabular}{c}
	\includegraphics[trim={0.5cm 0.7cm 0.5cm 0.5cm},width=\textwidth]{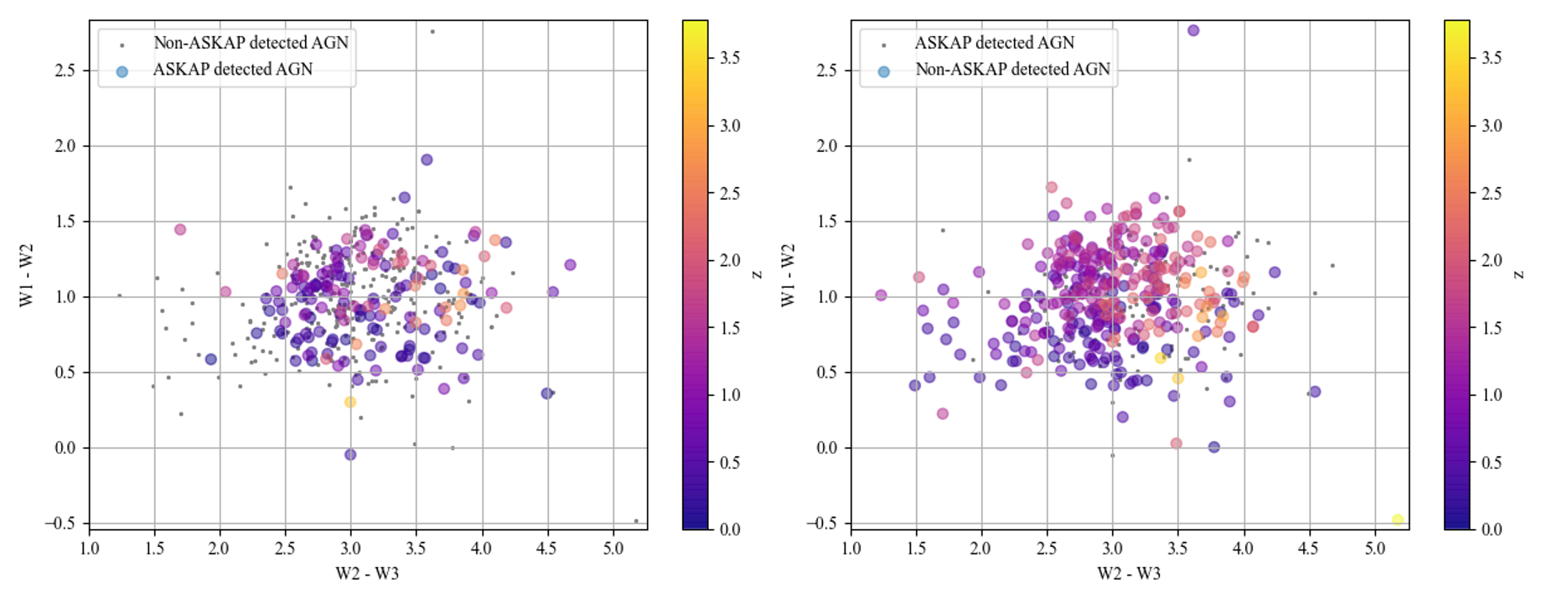}
	\end{tabular}
    \caption{(Left) AllWISE colour--colour diagram of ASKAP detected AGN, where the colour indicates redshift. Grey dots indicate the non-ASKAP detected AGN. (Right) AllWISE colour--colour diagram of non-ASKAP detected AGN, where the colour indicates redshift. Grey dots indicate the ASKAP detected AGN. Both the detected and non-detected AGN show a general gradient in redshift as a function of WISE colour, with exceptions.}
    \label{fig:specradio_CC_zcomp}
\end{figure*}

We compare the redshift distributions of ASKAP and non-ASKAP detected AGN using the same AllWISE colours as before and colouring based on redshift. (Figure \ref{fig:specradio_CC_zcomp}) shows that both the detected and non-detected AGN show a general gradient in redshift as a function of WISE colour, with a few outliers.

\begin{figure*}

	\includegraphics[width=\textwidth]{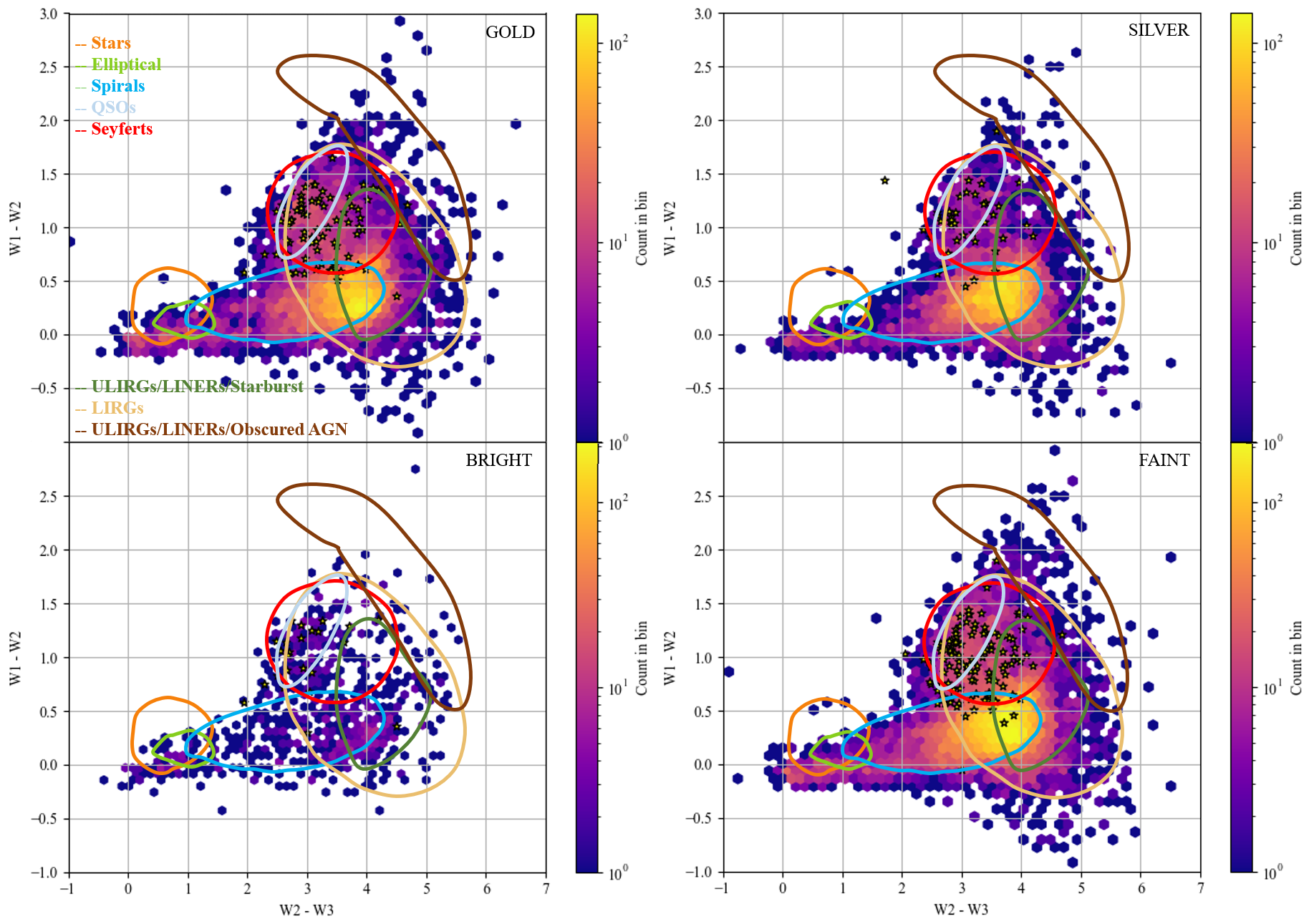}

    \caption{LMC AllWISE colour--colour diagram comparing the populations of the G\textsc{old} (top-left), S\textsc{ilver} (top-right), Bright (lower-left; $F_{888}>3$ mJy) and Faint (lower-right; $F_{888}<3$ mJy) source lists. The stars represent the spectroscopically observed AGN in their respective source list.  }
    \label{fig:LMC_WISEcolour_catcomp}
\end{figure*}

Figure \ref{fig:LMC_WISEcolour_catcomp} (top-left \& right) shows a WISE colour--colour diagram for the ASKAP detected sources and the known AGN in the G\textsc{old} and S\textsc{ilver} samples. Using the outlined regions for "QSOs" and "Spirals", in the G\textsc{old} list $\sim445$ out of 4852 ($\sim9.2$ per cent) are predicted to be AGN \citep[from AllWISE colours][]{Wright2010} and in the S\textsc{ilver} $\sim270$ out of 5050 ($\sim5.4$ per cent) are predicted to be AGN. As expected, there are more spirals compared to AGN in the S\textsc{ilver} list, since spirals tend to be star-forming and that is where the radio emission comes from, which tends to be fainter than the radio emission from an AGN and therefore a higher fraction of AGN are expected in the G\textsc{old} list. 

The upturn towards the fainter flux densities at $F_{888} \sim 3$ mJy can be seen in the overall flux distribution (see Figure \ref{fig:flux_hist}) representing the beginning of the faint galaxy population. We thus separated the catalogue into a bright sample ($F_{888}>3$ mJy) and a faint sample ($F_{888}<3$ mJy). Cross-matching this with the AllWISE catalogue provides 806 and 13,414 sources, respectively, with a maximum magnitude error in the W1, W2 and W3 bands of 0.2 mag. In Figure \ref{fig:LMC_WISEcolour_catcomp} (lower panels) we plot these two samples on a colour--colour diagram, which shows the distributions of the source types in the bright and faint samples, with regions of expected classifications annotated. From the colours shown in the plots the number of spirals in the faint population is greater than the number of AGN, whereas in the bright sample the ratio of spirals to AGN is lower. As expected the spiral galaxies are more prevalent in the faint population. Cross-matching with the spectroscopically observed AGN list gives 123 radio-faint AGN and 24 radio-bright AGN. The fraction of AGN is larger in the bright sample ($\sim3$ per cent) than in the faint sample ($\sim0.9$ per cent) as expected.

\section{Spectral indices}

We define the spectral index $\alpha$ by $F_\nu\propto\nu^\alpha$, where $F_\nu$ is the integrated flux density at frequency $\nu$. A flatter spectral index close to zero indicates free--free emission, and a steep negative spectral index, $\sim-0.7$, indicates synchrotron emission. Archival ATCA data at 1.384 GHz \citep{Hughes2007} cover a similar area as the ASKAP image, though with a larger beam size of 40$^{\prime\prime}$. We cross-matched the ASKAP LMC 888 MHz catalogue (G\textsc{old}, S\textsc{ilver} and B\textsc{ronze}) with the ATCA 1.384 GHz source list described in \cite{FilipovicTest} with a search radius of 5$^{\prime\prime}$. Limiting the sources to those with a $\leq$ 1 per cent error in ASKAP and ATCA integrated flux densities (determined by A\textsc{egean}) provides 1,869 sources.

\begin{figure}

	\includegraphics[trim={0.5cm 0 0.5cm 0 },width=\columnwidth]{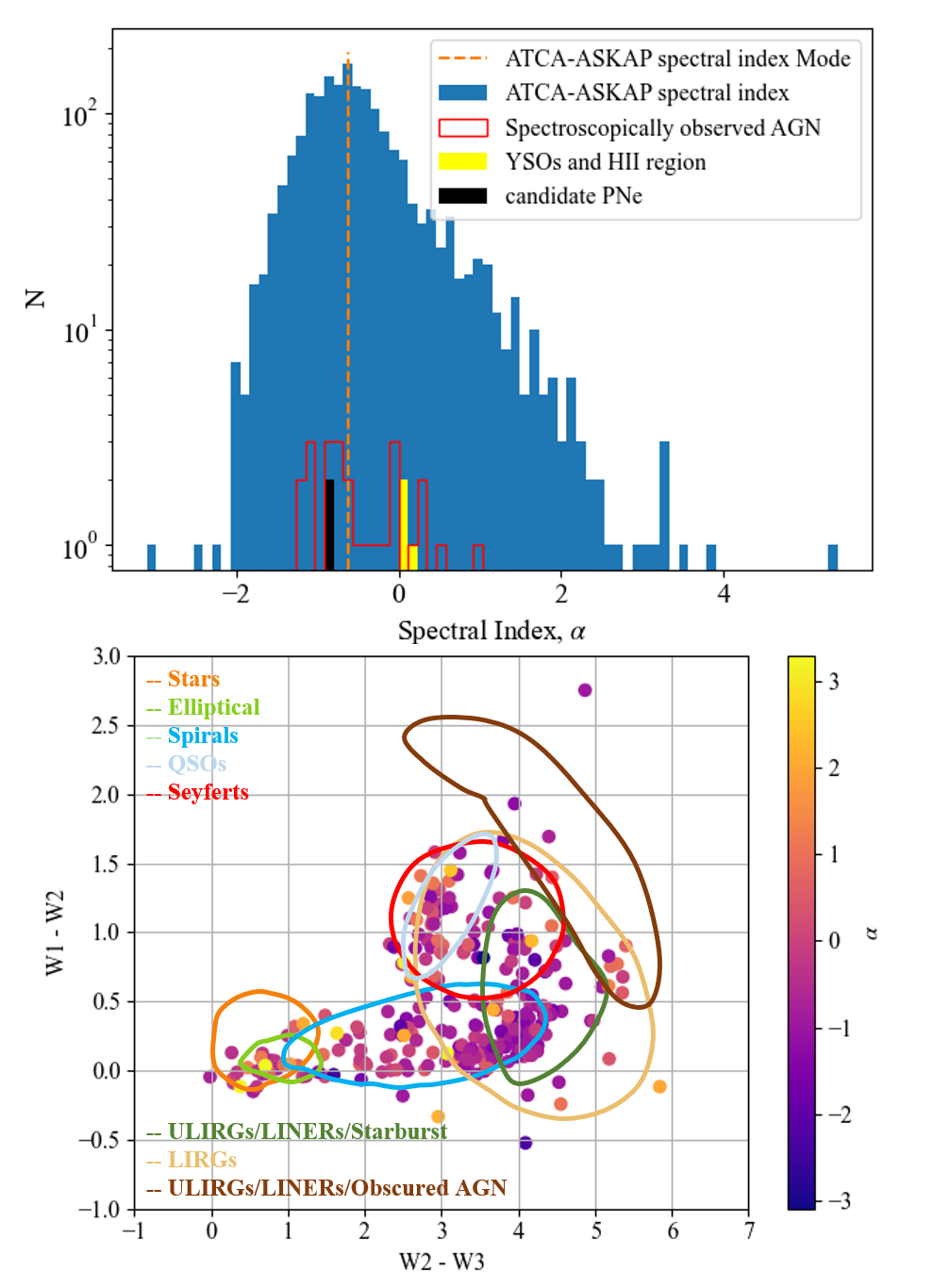}

    \caption{(Upper) Distribution of spectral indices calculated from ATCA (1.4 GHz) and ASKAP (888 MHz) integrated flux densities.  (Lower) WISE colour--colour diagram of 287 sources in ASKAP, ATCA and AllWISE.}
    \label{specind}
\end{figure}

The histogram of spectral indices (Figure \ref{specind} (upper)) is dominated by an extragalactic population of synchrotron sources ($\alpha<0$). This is confirmed by the small sample of 25 spectroscopically confirmed AGN -- it is not unusual for some AGN to show a flatter spectrum or even to peak at a few GHz. On the other hand, two YSOs (SAGE\,053054.2$-$683428.3 from Table 4, and [BE74]\,615 from Table 3) and an H\,{\sc ii} region ([RP2006]\,1933 from Table 3) have $\alpha\sim0$ consistent with their free--free emission spectrum. The two "candidate PNe" (SMP\,LMC\,32 and [RP2006]\,338, Table \ref{PNetable}) at $\alpha\sim-0.8$ are in fact among the few PN candidates that we had rejected on the basis of the large (in these cases $\sim9^{\prime\prime}$) distance from the radio position -- their synchrotron nature confirms that they are instead background sources.

We suspect that sources with $|\alpha|>3$ are the result of variability between the epochs at which the ATCA and ASKAP data were obtained. The most extreme case with $\alpha=5.3$, EMU EC J041753.5$-$731556 has a spectral index of $-0.69$ based on SUMSS instead of ASKAP, while EMU EC J054929.5$-$703545 with $\alpha=3.2$ has spectral index $-1.28$ when using SUMSS. The second-most extreme case, EMU EC J052029.2$-$680051 with $\alpha=3.9$ sits at the edge of H\,{\sc ii} nebula LHA\,120-N\,41 associated with emission-line star AL\,139 making the comparison with ATCA data problematic. The other six radio point sources with $|\alpha|>3$ generally have faint or no visible optical/IR counterparts in DSS, 2MASS and WISE. These could be extreme radio-variable sources at high redshift. It should be noted that at these frequencies there is the possibility of free--free absorption, which can rapidly absorb the power-law synchrotron emission spectrum towards lower radio frequencies \citep{Clemens2010}. This would cause the radio spectrum to turn-over/peak, which could lead to rapid changes in spectral index, such as those seen here.

\section{Conclusions}

We present the 888~MHz ASKAP EMU ESP radio continuum survey of the LMC.
Our findings can be summarised as follows:
\begin{itemize}
    \item This new ASKAP survey is a significant improvement (factor of $\sim$ 5 in the median RMS) compared to previous ATCA/MOST surveys of the LMC, and compared to the ASKAP survey of the SMC observed with a partial array (factor of $\sim$ 3 in the median RMS). The improvement in angular resolution has also allowed for greater detail of radio structure to be observed.
    \item We extracted 30,866, 22,080 and 1,666 sources at 888 MHz separated into G\textsc{old}, S\textsc{ilver} and B\textsc{ronze} source lists, respectively, with the majority of these sources detected above the 5$\sigma$ threshold. 
    \item We found 114 out of 1,334 PNe listed in Simbad to have a counterpart in ASKAP within 10 arcsec. Only 37 of these detections are likely true PNe, with the remaining sources being emission-line stars, candidate post-AGB objects, YSOs and compact H\,{\sc ii} regions.
    \item YSOs detected with {\it Herschel} \citep{Oliveira2019} were discovered to show a strong correlation between integrated radio flux densities and gas temperature, with those with $T > 700$ K detected in radio and those with $T < 700$ K not detected in radio. This suggests a strong relation between the line emission from the photo-dissociation regions at the interfaces of neutral and ionized gas, and the free--free emission from the ionized regions surrounding the nascent, presumed massive O-type stars.
    \item Of 46 SNe listed in Simbad, most associated radio detections could not be unequivocally ascribed to the SN itself and are more likely emission from a background host galaxy. Only SN\,1978A and Nova\,LMC\,1988\,b are securely detected with ASKAP.
    \item Exploring the GLEAM 4-Jy sample \citep{White2020b,White2020a} in the ASKAP field of the LMC revealed six GLEAM sources, showcasing that the ASKAP is an improvement in angular resolution (from $\sim$ 2$^{\prime}$ to $\leq$ 14$^{\prime\prime}$), allowing for radio sources to be better traced to their optical/IR counterparts and revealing morphology associated with jets and likely interaction with the intra-cluster medium.
    \item Cross-matching the new LMC ASKAP source catalogue with AllWISE reveals 14,333 sources with photometry errors $<$ 0.2 mag. Comparing the radio and non-radio sources reveals that the majority of the radio sources are extragalactic.
    \item Cross-matching the new LMC ASKAP source catalogue with the miliquas catalogue \citep{Flesch2019pap,Flesch2019cat} reveals 190 radio detections out of 657 spectroscopically confirmed AGN. More than 92 per cent of these are newly detected here.
    \item A higher fraction of AGN is found within the bright population  ($F_{888}>3$ mJy) than within the faint population ($F_{888}<3$ mJy). As expected there is a higher fraction of spirals in the faint population compared to the bright population.
    \item Cross-matching the new ASKAP 888 MHz data with archival ATCA 1.4 GHz data reveals a spectral index distribution peaking at $\alpha=-0.6$ (where $F_\nu\propto\nu^\alpha$). This corroborates that the majority of sources are extragalactic synchrotron emitters. Extreme spectral indices ($|\alpha|>3$) can be explained by variability.
\end{itemize}

Future work will include more comprehensive analyses of the planetary nebul{\ae} and supernova remnants, as well as the radio variability. We are employing machine learning to classify a combined multi-wavelength data set and study the radio--IR relation for galaxies and quasars. The background population seen in the ASKAP image shows the potential for large-scale Faraday rotation and H\,{\sc i} absorption measurements throughout the LMC.

\section{Data Availability}
The catalogue and PNe table created in this work are available as supplementary material to this article. The image is available to the public through CASDA\footnote{https://research.csiro.au/casda/}. The catalogue, image and the PNe table will be made available on the CDS\footnote{https://cds.u-strasbg.fr/} website when the paper is published. 

\section*{Acknowledgements}


We thank the anonymous referee for their feedback, which helped improve the paper. The Australian SKA Pathfinder is part of the Australia Telescope National Facility, which is managed by CSIRO. Operation of ASKAP is funded by the Australian Government with support from the National Collaborative Research Infrastructure Strategy. ASKAP uses the resources of the Pawsey Supercomputing Centre. Establishment of ASKAP, the Murchison Radio-astronomy Observatory and the Pawsey Supercomputing Centre are initiatives of the Australian Government, with support from the Government of Western Australia and the Science and Industry Endowment Fund. We acknowledge the Wajarri Yamatji people as the traditional owners of the Observatory site.

CMP is supported by an STFC studentship awarded to Keele University. M.J.M.~acknowledges the support of the National Science Centre, Poland through SONATA BIS grant 2018/30/E/ST9/00208. Partial support for L.R. comes from US National Science Foundation grant AST17-14205 to the University of Minnesota. S.A. acknowledges financial support from the Centre for Research \& Development in Mathematics and Applications (CIDMA) strategic project UID/MAT/04106/2019 and from Enabling Green Escience for the Square Kilometre Array Research Infrastructure (ENGAGESKA), POCI-01-0145-FEDER-022217, funded by the Programa Operacional Competitividade e Internacionalização (COMPETE 2020) and FCT, Portugal. This research  made use of the SIMBAD database, operated at CDS, Strasbourg, France, and the NASA/IPAC Extragalactic Database (NED) which is operated by the Jet Propulsion Laboratory, California Institute of Technology, under contract with the National Aeronautics and Space Administration.

This work has made use of Python software packages: aegean\footnote{https://github.com/PaulHancock/Aegean}, astropy \footnote{http://www.astropy.org}, numpy\footnote{http://numpy.org}, pandas\footnote{http://pandas.pydata.org} and matplotlib\footnote{http://matplotlib.org}.




\bibliographystyle{mnras}
\bibliography{ref} 




\appendix



\vspace{10mm}
\textit{
\hspace{1mm}\\
$^{2}$Western Sydney University, Locked Bag 1797, Penrith South DC, NSW 2751, Australia\\
$^{3}$Depto.\ de Astronom\'ia, DCNE, Universidad de Guanajuato, Cj\'on.\ de Jalisco s/n, Col.\ Valenciana, Guanajuato, CP 36023, Gto., M\'{e}xico\\
$^{4}$Max-Planck-Institut f{\"u}r extraterrestrische Physik, Gie{\ss}enbachstra{\ss}e 1, D-85748 Garching bei M\"unchen, Germany\\
$^{5}$Dominion Radio Astrophysical Observatory, Herzberg Programs in Astronomy and Astrophysics, National Research Council Canada, P.O.\ Box 248, Penticton, BC V2A 6J9, Canada\\
$^{6}$Australia Telescope National Facility, CSIRO Astronomy and Space Science, P.O.\ Box 76, Epping, NSW 1710, Australia\\
$^{7}$Minnesota Institute for Astrophysics, School of Physics and Astronomy, University of Minnesota, 116 Church Street SE, Minneapolis, MN 55455, USA\\
$^{8}$Department of Physics and Electronics, Rhodes University, P.O.\ Box 94, Grahamstown, 6140, South Africa\\
$^{9}$European Southern Observatory, Karl-Schwarzschild-Strasse 2, Garching bei M{\"u}nchen, 85748, Germany\\
$^{10}$Centre for Research and Development in Mathematics and Applications (CIDMA), Campus de Santiago, 3810-183 Aveiro, Portugal\\
$^{11}$Research Department, Plasmas with Complex Interactions, Ruhr-Universit\"{a}t Bochum, 44780 Bochum, Germany\\
$^{12}$Astronomisches Institut, Ruhr-Universit{\"a}t Bochum, 44780 Bochum, Germany\\
$^{13}$The Inter-University Institute for Data Intensive Astronomy (IDIA), Department of Astronomy, University of Cape Town, Rondebosch 7701, South Africa\\
$^{14}$Australian Astronomical Optics, Macquarie University, 105 Delhi Rd., North Ryde, NSW 2113, Australia\\
$^{15}$CSIRO Astronomy and Space Science, Radiophysics Laboratory, Marsfield, NSW, Australia\\
$^{16}$School of Cosmic Physics, Dublin Institute for Advanced Studies, 31 Fitzwilliam Place, Dublin 2, Ireland\\
$^{17}$Department of Physics and Astronomy, University of Calgary, University of Calgary, Calgary, Alberta, T2N 1N4, Canada\\
$^{18}$Observatoire Astronomique de Strasbourg, Universit\'{e} de Strasbourg, CNRS, 11 rue de l’Universit\'{e}, F-67000 Strasbourg, France\\
$^{19}$National Radio Astronomy Observatory, 1003 Lopezville Rd., Socorro, NM 87801, USA\\
$^{20}$Astronomical Observatory Institute, Faculty of Physics, Adam Mickiewicz University, ul. S\l{}oneczna 36, 60-286 Pozna\'{n}, Poland\\
$^{21}$National Astronomical Observatory of Japan, Mitaka, Tokyo 181-8588, Japan\\
$^{22}$Department of Physics, Nagoya University, Furo-cho, Chikusa-ku, Nagoya 464-8601, Japan\\
$^{23}$Institute for Advanced Research, Nagoya University, Furo-cho, Chikusa-ku, Nagoya 464-8601, Japan\\
$^{24}$Remeis Observatory and ECAP, Universit{\"a}t Erlangen-N{\"u}rnberg, Sternwartstra{\ss}e 7, D-96049 Bamberg, Germany\\
$^{25}$International Centre for Radio Astronomy Research, University of Western Australia, 35 Stirling Hwy, Crawley, WA 6009, Australia\\
$^{26}$ARC Centre of Excellence for All Sky Astrophysics in 3 Dimensions (ASTRO 3D)\\
$^{27}$Th\"{u}ringer Landessternwarte, Sternwarte 5, 07778 Tautenburg, Germany
}

\bsp	
\label{lastpage}
\end{document}